\documentclass[10pt,epsfig]{article} 
\pdfoutput=1
\usepackage{pstricks}
\usepackage{enumerate}
\usepackage{float}
\usepackage{subfig}
\usepackage{color}
\usepackage{slashed}
\usepackage{amssymb, amsmath,mathrsfs}
\usepackage{graphicx}
\usepackage{multirow}
\usepackage{cite}
\usepackage{chngcntr}
\usepackage[colorlinks=true,
            linkcolor=red,
            urlcolor=blue,
            citecolor=blue]{hyperref}
\usepackage{dcolumn}
\usepackage{bm}
\usepackage{color}
\long\def\rpl#1!!#2!!{\textcolor{red}{#1} \textcolor{blue}{#2}}
\def\baselinestretch{1.3}
\newcommand{\ba}{\begin{array}}
\newcommand{\ea}{\end{array}}
\newcommand{\bd}{\begin{displaymath}}
\newcommand{\ed}{\end{displaymath}}
\newcommand{\besub}{\begin{subequations}}
\newcommand{\eesub}{\end{subequations}}
\newcommand{\be}{\begin{equation}}
\newcommand{\ee}{\end{equation}}
\newcommand{\bea}{\begin{eqnarray}}
\newcommand{\eea}{\end{eqnarray}}
\newcommand{\no}{\nonumber\\}

\def\a{\alpha}

\def\b{\beta}

\def\l{\lambda}

\def\L{\Lambda}

\def\q2 {q^2}

\def\bt{\begin{table}}
\def\et{\end{table}}

\catcode`@=11 % This allows us to modify PLAIN macros.
\def \gsim{\mathrel{\mathpalette\@versim>}}
\def \lsim{\mathrel{\mathpalette\@versim<}}
\def \@versim#1#2{\lower0.4ex\vbox{\baselineskip\z@skip\lineskip\z@skip
     \lineskiplimit\z@\ialign{$\m@th#1\hfil##\hfil$
     \crcr#2\crcr\sim\crcr}}}
\catcode`@=12 % at signs are no longer letters

\allowdisplaybreaks
\begin{document}

%THE TEXT STARTS HERE
\begin{flushright}
{HRI-RECAPP-2016-004}
\end{flushright}

\begin{center}

{\large \textbf {High-scale validity of a two Higgs doublet scenario: metastability included.}}\\[15mm]

Nabarun Chakrabarty$^{\dagger}$\footnote{nabarunc@hri.res.in} and Biswarup Mukhopadhyaya$^{\star}$\footnote{biswarup@hri.res.in}  \\
$^{\dagger}${\em Regional Centre for Accelerator-based Particle Physics \\
     Harish-Chandra Research Institute\\
 Chhatnag Road, Jhunsi, Allahabad - 211 019, India}\\[5mm] 

\end{center}

\begin{abstract} 

We make an attempt to identify regions in a Type II Two-Higgs Doublet Model, which correspond
to a metastable electroweak vacuum with lifetime larger than the age of the universe. We analyse
scenarios which retain perturbative unitarity up to Grand unification and Planck scales. 
Each point in the parameter space is restricted using Data from
the Large Hadron Collider (LHC) as well as flavor and precision electroweak constraints. We find that
substantial regions of the parameter space are thus identified as corresponding to metastability,
which compliment the allowed regions for absolute stability, for top quark mass at the high as well
as low end of its currently allowed range. Thus, a two-Higgs doublet scenario with
the electroweak vacuum, either stable or metastable, can sail through all the way up to the Planck
scale without facing any contradictions. 

\end{abstract}

\newpage
\setcounter{footnote}{0}

\def\baselinestretch{1.5}
\counterwithin{equation}{section}
%==========================================================================
%==========================================================================
\section{Introduction}\label{Intro}

With the observation of a scalar resonance around 125 GeV at the LHC\cite{Aad:2012tfa,Chatrchyan:2012xdj}, and hence its identication with a 
Higgs boson, the particle spectrum of the Standard Model (SM) appears to be complete. However, 
issues ranging from the existence of Dark Matter (DM) to the pattern of neutrino mass continue to suggest physics beyond the
SM. While the quest for such new physics remains on, a rather pertinent question to ask is 
whether the SM by itself can ensure vacuum stability at scales above that of electroweak
symmetry breaking. This is because the
Higgs quartic coupling evolving via SM interaction alone tends to turn negative in between the Electroweak (EW)
and Planck scales, thereby making the scalar potential unbounded from below. This exact location of this instability scale crucially depends on the pole masses of the top quark and the Higgs. 
A recent next-to-next-to-leading order (NNLO) study \cite{Degrassi:2012ry,Buttazzo:2013uya}
finds that absolute stability up to the Planck scale requires\cite{Degrassi:2012ry}
\begin{eqnarray}
M_h [\rm GeV] > 129.4 + 1.4(\frac{M_t[\rm GeV]-173.1}{0.7})-0.5(\frac{\alpha_s(M_Z)-0.1184}{0.0007})\pm1.0_{th} 
\end{eqnarray}
The updated measurements of the Higgs and top quark masses [REF] hint towards
a \emph{metastable} vacuum scenario. Such a scenario leads to an additional minimum
in the potential, which is deeper than the electroweak vacuum\cite{Isidori2001387}. However the lifetime of 
the latter is less than the age of the universe, thus enabling the present day vacuum
to be consistent with the well tested electroweak theory.

In general, vacuum instability can be 
alleviated by introducing additional bosonic degrees of freedom, which can offset the downward
evolution of the quartic coupling of the SM. Such a possibility has indeed been explored in the context of various non-minimal
Higgs sectors. One example would be the case of the celebrated Two-Higgs Doublet Models (2HDMs).
Different types of 2HDM offer interesting phenomenology at present and future colliders and, are consistent with flavor physics constraints, and are part and parcel of supersymmetric theories. In one of our earlier works, we showed in context 
of a Type-II 2HDM that the EW
vacuum  can be rendered stable till the Planck scale even for a top pole mass at the high
end of the allowed band~\cite{Agashe:2014kda}. Moreover, this can be achieved without running into conflict
with perturbativity or unitarity at high scales.

A 2HDM is set has one more vital feature. The Yukawa couplings of the SM fermions can
be different compared to the SM values, in a 2HDM. The role of the coupling for the same $M_t$ can thus produce a different effect to the evolution to the quartics. A stable
vacuum till the Planck scale is achieved with specific combinations of the boundary conditions,
i.e, given in terms of the model parameters.
This raises the question, \emph{is there a possible metastable vacuum in a 2HDM? Can such a balance between the bosonic and fermionic effects be struck that indeed leads to an 
additional minimum of the scalar potential, while prolonging the lifetime of the EW vacuum to a safe level? This is the question we precisely seek to answer
in this work.}

Metastability in context of certain non-minimal Higgs scenarii has been looked into in the recent times and along with the attempts  
%in recent times, see for example in REF. In REF, the authors have investigated the possibility
of embedding successful cold dark matter canditates in the parameter space that leads to
a stable/metastable vacuum. These studies however confront \emph{inert} scalars that do not
mix with the SM Higgs owing to some discrete symmetry, and moreover, the Higgs-top coupling
also remains unchanged with respect to the SM. The implications could be different in a
2HDM where EWSB is triggered when both the doublets receive vacuum expectation values (vev).
Two possibilities open thus up: (a) The scalar potential could furnish 
additional neutral minima around the TeV
scale ballpark in the slice spanned by the neutral fields in the two doublets, and, (b) Additional
minimum can appear when the scalar potential is improved by Renormalisation Group (RG) effects.
In the process of looking for such possibilities, we take into account miscellaneous
constraints coming from Higgs signal strengths, flavor issues and electroweak precision observables.

The paper has the following plan. In Sec.~\ref{Model}, we review the salient features of the
2HDMs. Sec.~\ref{Probability} is dedicated to a discussion on how a metastable vacuum can arise, and on the 
completion of its lifetime. We also present an outline of the tunnelling probability computation in the same.
Sec.~\ref{Numan} presents an overall strategy on how to look for a metastable vacua, and, also an 
account of the various experimental and theoretical constraints taken while doing so. The 
numerical results are highlighted in Sec.~\ref{Results} and finally the study is concluded in Sec.~\ref{Conclusions}.

\section{Model features.}\label{Model}
\subsection{Scalar potential}
In the present work, we consider the most general renormalizable
scalar potential for two doublets $\Phi_1$ and $\Phi_2$, each having
hypercharge $(+1)$,
\bea
V(\Phi_1,\Phi_2) &=&
m^2_{11}\, \Phi_1^\dagger \Phi_1
+ m^2_{22}\, \Phi_2^\dagger \Phi_2 -
 m^2_{12}\, \left(\Phi_1^\dagger \Phi_2 + \Phi_2^\dagger \Phi_1\right)
+ \frac{\lambda_1}{2} \left( \Phi_1^\dagger \Phi_1 \right)^2
+ \frac{\lambda_2}{2} \left( \Phi_2^\dagger \Phi_2 \right)^2
\no & &
+ \lambda_3\, \Phi_1^\dagger \Phi_1\, \Phi_2^\dagger \Phi_2
+ \lambda_4\, \Phi_1^\dagger \Phi_2\, \Phi_2^\dagger \Phi_1
+ \frac{\lambda_5}{2} \left[
\left( \Phi_1^\dagger\Phi_2 \right)^2
+ \left( \Phi_2^\dagger\Phi_1 \right)^2 \right]
\no & &
+\lambda_6\, \Phi_1^\dagger \Phi_1\, \left(\Phi_1^\dagger\Phi_2 + \Phi_2^\dagger\Phi_1\right) + \lambda_7\, \Phi_2^\dagger \Phi_2\, \left(\Phi_1^\dagger\Phi_2 + \Phi_2^\dagger\Phi_1\right).
\label{pot}
\eea

 This scenario in general has the
possibility of CP-violation in the scalar sector, through the phases in  $m_{12}$, $\lambda_5$, $\lambda_{6}$ and $\lambda_7$. We choose  $m_{12}$ to be real here,
moreover the terms proportional 
to $\l_6$ and $\l_7$ have been neglected in the present study.

In a general two-Higgs-doublet model (2HDM), a particular fermion can
couple to both $\Phi_1$ and $\Phi_2$. However this would lead to the
flavor changing neutral currents (FCNC) 
at the tree level\footnote{In context of a typical flavour changing scenario, it has been shown recently that the FCNCs are stable under RG evolution mostly.}.
One way to avoid such FCNC is to impose a $\mathbb{Z}_2$ symmetry, such as one
that demands invariance under $\Phi_1 \to -\Phi_1$ and $\Phi_2 \to
\Phi_2$. This type of symmetry puts restrictions on
the scalar potential. The $\mathbb{Z}_2$ symmetry is \emph{exact} as
long as $m_{12}$, $\lambda_{6}$ and $\lambda_7$ vanish, when the scalar
sector also becomes CP-conserving. The symmetry
is said to be broken \emph{softly} if it is violated in the quadratic
terms only, i.e., in the limit where $\lambda_{6}$ and $\lambda_7$
vanish but $m_{12}$ does not. Finally, a \emph{hard} breaking of the
$\mathbb{Z}_2$ symmetry is realized when it is broken by the quartic
terms as well. Thus in this case, $m_{12}$, $\lambda_{6}$ and
$\lambda_7$ all are non-vanishing in general.
 
We focus on a particular scheme
of coupling fermions to the doublets. This scheme is known in
the literature as the \emph{Type-II} 2HDM, where the down type quarks
and the charged leptons couple to $\Phi_1$ and the up type quarks, to
$\Phi_2$\cite{Branco:2011iw}. This is ensured through the discrete symmetry $\Phi_{1}\to -\Phi_{1}$ and $\psi_{R}^{i}\to -\psi_{R}^{i}$, 
where $\psi$ is charged leptons or down type quarks and $i$ represents the generation index. 

Minimization of the scalar potential in Eqn.~\ref{pot} leads to 
\be
\langle \Phi_1 \rangle
= \left( \begin{array}{c} 0 \\ \displaystyle{\frac{v_1}{\sqrt{2}}} \end{array} \right),
\quad
\langle \Phi_2 \rangle
= \left( \begin{array}{c} 0 \\ \displaystyle{\frac{v_2}{\sqrt{2}}} \end{array} \right),
\ee  
\noindent
where the vacuum expectation values (vev) are often expressed in terms of 
the $M_Z$ and the ratio
\be
\tan \beta = \frac{v_2}{v_1}\;. 
\ee
We parametrise the doublets in the following fashion,
\be
\Phi_{i} = \frac{1}{\sqrt{2}} \begin{pmatrix}
\sqrt{2} w_i^{+} \\
v_i + h_i + i z_i
\end{pmatrix}~ \rm{for}~\textit{i} = 1, 2.
\label{e:doublet}
\ee
Since the basis used in $V(\Phi_1,\Phi_2)$ allows mixing between the two
doublets, the physical states are obtained by diagonalising the charged and neutral scalar mass
matrices. There are then altogether eight
mass eigenstates, three of which become the longitudinal components
of the $W^{\pm}$ and $Z$ gauge bosons.  Of the remaining five, there is a 
mutually conjugate pair of charged scalars ($H^{\pm}$), two neutral CP even
scalars ($H, h$) and a neutral pseudoscalar ($A$), given there is no
CP-violation. Otherwise, a further mixing between ($H, h$) and $A$ becomes unavoidable. 
The compositions of the mass eigenstates $H$ and $h$ indeed depend on the mixing angle $\alpha$.

The quartic couplings are conveniently expressed in terms of the physical masses
and mixing angles as,
\besub
\bea
\label{e:masq}
\l_1 &=& \frac{1}{v^2 c^2_\b}~\Big(c^2_\a m^2_H + v^2 s^2_\a m^2_h - m^2_{12}\frac{s_\b}{c_\b}\Big)\\
\l_2 &=& \frac{1}{v^2 s^2_\b}~\Big(s^2_\a m^2_H + v^2 c^2_\a m^2_h - m^2_{12}\frac{c_\b}{s_\b}\Big)\\
\l_3 &=& \frac{2 m^2_{H^+}}{v^2} + \frac{s_{2\a}}{v^2 s_{2\b}}~(m^2_H - m^2_h) - \frac{m^2_{12}}{v^2 s_\b c_\b}\\
\l_4 &=& \frac{1}{v^2}~(m^2_A - 2 m^2_{H^+}) + \frac{m^2_{12}}{v^2 s_\b c_\b}\\
\l_5 &=& \frac{m^2_{12}}{v^2 s_\b c_\b} - \frac{m^2_A}{v^2}
\eea
\label{e:Couplings}
\eesub

\section{The computation of tunneling probability.}\label{Probability}

The existence of a large number of scalar degrees of freedom makes the
vacuum landscape of a 2HDM more elaborate and intriguing compared to the SM. 
%Apart from the canonical normal vacua, i.e, when only the CP even even fields pick
%up a non-zeo vacuum expectation value, charge breaking and CP breaking vacua could 
%also arise in a generic 2HDM. 
%For the normal vacua depicted in Eqn, the tadpole conditions read,
Here we are confining ourselves to the situation when the vacuum breaks neither
electric charge nor CP. Under such circumstances, the EWSB conditions appear as,
\bea
m^2_{11} v_1 &=& m^2_{12} v_2 - \frac{1}{2}\l_1 v_1^3  - \frac{1}{2}(\l_3 - \l_4 + \l_5) v_1 v_2^2 \\
m^2_{22} v_2 &=& m^2_{12} v_1 - \frac{1}{2}\l_2 v_2^3  - \frac{1}{2}(\l_3 - \l_4 + \l_5) v_2 v_1^2
\eea
It has been reported in~\cite{Barroso:2013awa} that the above conditions can lead to several solutions, and
at most two non-degenerte minima. In other words, apart from the EW minimum in which
the universe currently resides ($v_1^2 + v_2^2 = 246 ~\text{GeV}^2$, named $N$),
 there exists another minimum somewhere around ($v_1^2 + v_2^2 \neq 246 ~\text{GeV}^2$, named $N^{\prime}$). Ref. finds the difference of depths of the tree level scalar potential
at the two minima to be,
\be
V_{N^{\prime}} - V_{N} = \frac{m^2_{12}}{4 v_1 v_2}~\Big(1 - \frac{v_1 v_2}{v_1^{\prime}v_2^{\prime}}\Big)^2~(v_1v_2^{\prime} - v_2v_1^{\prime})^2
\ee
Thus there exists the tantalizing possibility that the 2HDM offers such parameter points 
for which a neighbouring vacuum could actually be deeper than the one which corresponds to the
observed $W-$ and $Z-$ boson masses. 
The EW minimum then loses its status as the global munimum and has been termed
the \emph{panic vacuum} in \cite{Barroso:2013awa}.
In those cases, computing the lifetime of tunnelling to the non-EW minimum
from the EW one becomes the pertinent task. If the tunnelling lifetime turn out to be
higher than the age
of the universe, the non-EW minimum cannot be ruled out. 
However, thanks to the data from the LHC on Higgs signal strengths, the model points
admitting $V_{N^{\prime}} - V_{N}$ $<$ 0 are more or less ruled out\cite{Chakraborty:2015raa,Barroso:2013awa}. However, a new landscape of vacua can still 
open up if one 
investigates the \emph{renormalisation-group improved} effective potential in 
place of the bare tree-level one. In the context of the SM, it can be understood
as follows: The SM quartic coupling turns negative at some energy scale $10^{8-11}$ GeV
(exact location of the scale depends on the choice of the initial conditions), after 
which it again starts rising owing to the bosonic effects counterbalancing the
negative top-Yukawa drag. The fallout of this is the emergence of a new minimum beyond
the scale where the quartic coupling first becomes negative. It should be noted that
the direction of the EW vacuum uniquely decides the direction in which the high-scale
vacuum is formed.

In a 2HDM, on the other hand, one has to handle the 
additional complication of having a higher number of field directions. In addition, the
effects of the various interaction terms make it imperative to incorporate 
the effects of radiative corrections induced by the 2HDM. Therefore, we choose to 
analyse the one-loop corrected effective potential in place of the tree level potential.
One thus writes
\bea
V_{eff}(h_1,h_2) &=& V_{tree}(h_1,h_2) + V_{1loop}(h_1,h_2)
\eea
Here, $V_{tree}(h_1,h_2)$ and $V_{1loop}(h_1,h_2)$ denote the tree level and 1-loop
parts of the effective potential, calculated along the $h_1 - h_2$ subspace. 

For example, 
, the tree level potential reads
\bea
V_{tree}(h_1,h_2) &=& \frac{1}{2} m^2_{11} h^2_1 + \frac{1}{2} m^2_{22} h^2_2 + m^2_{12} h_1 h_2 + \frac{\l_1}{8} h^4_1 + \frac{\l_2}{8} h^4_2 + \frac{\l_3 + \l_4 + \l_5}{4} h^2_1 h^2_2
\eea

 In the 
($h_1,h_2$) plane, it has
the following expression,
\bea
V_{1loop}(h_1,h_2) &=&
\frac{1}{64 \pi^2} \sum_{i} n_i M^4_i(h_1,h_2)\Bigg[\text{ln}\Big(\frac{M^4_i(h_1,h_2)}{\mu^2}\Big) - c_i \Bigg] 
\label{pot}
\eea
Where $n_i$ refers to the number of degrees of freedom for the ith field and $c_i$ are 
constants whose values depend on the regularization scheme adopted. To list the 
constants explicitly, $n_W = 6$, $n_Z = 3$ , $n_t = -12$, $n_h = 1$; and 
$c_W = \frac{5}{6}$, $c_Z = \frac{5}{6}$ , $c_t = \frac{3}{2}$, $c_h = \frac{3}{2}$.
Moreover $\mu$ refers to the renormalization scale emerging as an artifact
of dimensional regularization. $M^2_i(h_1,h_2)$ represent the scale dependent mass squared.

Studies on the high-scale validity of a 2HDM in the past were mostly confined to investigating
absolute stability\cite{Chakrabarty:2014aya,Das:2015mwa,Eberhardt:2014kaa,Chakrabarty:2015yia}
Some studies connecting higher dimensional operators to Higgs metastability have occured in the past\cite{Branchina:2013jra,PhysRevD.91.013003}.
The main theme of this work is to investigate possible high scale vacua in the context of 2HDM using the prescription suggested by Coleman. $V_{eff}(h_1,h_2)$ depends on two variables, and hence,
determining a classical solution interpolating the two vacua, even numerically, becomes an extremely challenging task. Furthermore, a generic classical path may not qualify as a "bounce"~\cite{PhysRevD.15.2929,PhysRevD.16.1762}, i.e, it
might not pass through the top of a barrier separating two vacua. Coleman's presciption does not apply in such a case. However, one can alway choose to look for additional minima along a particular ray in the $h_1 - h_2$ plane. Under this approximation, the effective potential is reduced to a function of a single variable again(that particular linear combination of $h_1$ and $h_2$).
In models such as Type I or Type II 2HDM, the $Z_2$ symmetry of the Yukawa interactions implies that the top quark always couples to $\Phi_2$. Thus it is
the coupling $\l_2$ that experiences the maximum downward pull due to the Yukawa interactions and can consequently can turn
negative at high scales in spite of starting with a positive value at the input scale. It therefore makes
sense to look for additional minima in the $h_2$ direction only. This approach is similar to what
\cite{Khan:2015ipa} opts in context of an inert doublet model.
 
We study the behaviour of the $V_{eff}(h_1,h_2)$ in the limit where $h_1\simeq v$ and $h_2$ $>>$ $h_1$, $m_{12}$.
In this limit,  
the squared masses have the following simplified expressions: 
\bea
m^2_{H_1}(h_2) &\simeq& \frac{1}{2}(\l_3 + \l_4 + \l_5)h^2_2 \\
m^2_{H_2}(h_2) &\simeq& \frac{3}{2}\l_2 h^2_2 \\
m^2_{A_1}(h_2) &\simeq& \frac{1}{2}(\l_3 + \l_4 - \l_5)h^2_2 \\
m^2_{A_2}(h_2) &\simeq& \frac{1}{2}\l_2 h^2_2 \\
m^2_{H^+_1}(h_2) &\simeq& \frac{1}{2}\l_3 h^2_2 \\
m^2_{H^+_2}(h_2) &\simeq& \frac{1}{2}\l_2 h^2_2 \\
m^2_{t}(h_2) &\simeq& \frac{1}{2}y^2_t h^2_2 \\
m^2_{W}(h_2) &\simeq& \frac{1}{4}g^2 h^2_2 \\
m^2_{Z}(h_2) &\simeq& \frac{1}{4}(g^2 + {g^{\prime}}^2) h^2_2
\eea

%We make the passing remark that whenever a field dependent mass is negative due to
%a particular choice of the quartics, the effective potential develops an imaginary
%part. However, the high-scale fate of the potential is solely dictated by the real part
%and thus, we opt to work with the real part only whenever so it occurs. 

All running couplings are evaluated at the scale $\mu \simeq h_2$.
Under all these approximations, the real part of one-loop corrected potential takes the form,
\bea
V_{eff}(h_2) &\simeq& \frac{\l^{eff}_2}{8}h^4_2 \\
\text{where,}\\ \nonumber
\l^{eff}_2(h_2) &\simeq& \l_2(h_2) + \frac{1}{64 \pi^2}\Bigg[2(\l_3 + \l_4 + \l_5)^2 \Big(\text{ln}\frac{\l_3 + \l_4 + \l_5}{2} - \frac{3}{2}\Big) + 18 \l_2^2 \Big(\text{ln}\frac{3\l_2}{2} - \frac{3}{2}\Big)\\ \nonumber
&&
2(\l_3 + \l_4 - \l_5)^2 \Big(\text{ln}\frac{\l_3 + \l_4 - \l_5}{2} - \frac{3}{2}\Big) + 2 \l_3^2 \Big(\text{ln}\frac{\l_3}{2} - \frac{3}{2}\Big) + 6 \l_2^2 \Big(\text{ln}\frac{\l_3}{2} - \frac{3}{2}\Big)\\ \nonumber
&&
3 g^4\Big(\text{ln}\frac{g^2}{4} - \frac{5}{6}\Big) + \frac{3}{2} (g^2 + {g^{\prime}}^2)^2\Big(\text{ln}\frac{g^2 + {g^{\prime}}^2}{4} - \frac{5}{6}\Big) - 24 y_t^4\Big(\text{ln}\frac{y_t^2}{2} - \frac{3}{2}\Big)\Big]
\eea

Where the term in square brackets refers to the
finite correction generated by the Coleman Weinberg mechanism. We find that highly sub-dominant in our calculations.

The probability of tunneling to the deeper vacuum is given by,
\be
p = {T_U^4}\mu^4 ~e^{-\frac{8\pi^2}{3 |\l_2^{eff}|}}
\ee
Here $\mu$ refers to the scale where the probability is maximized, and, it turns out that $\frac{d \l_2}{d \text{log(Q)}}$ = 0  at $Q = \mu$.
Using $T_U \simeq 10^{10}$ yr and requiring that the vacuum tunnelling lifetime is always higher than the lifetime on the universe tantamounts to having the following condition~\cite{Isidori2001387},
\be
\l_2^{eff}(h_{2}) \geq \frac{-0.065}{1 - 0.01 \text{~ln(a)}}
\ee

where $a = \frac{v}{\mu}$. 
It may be noted that we have accepted $\l_2$ turning negative in the $h_2$ direction as the
sole condition for the loss of stability of the EW vacuum. There is in general an extended set of conditions for stability in a 2HDM \cite{Branco:2011iw}. However, one can easily verify that the remaining conditions
for stability in a 2HDM are violated, if at all, at low scale itself. Such violation, on the other 
hand, leads to the disappearance of the EW minimum as a whole. This cannot be a situation
appropriate for metastability, and therefore the consitions other than $\l_2 < 0$ need not be
used as signs for loss of stability.

\section{A metastable vacuum and the 2HDM parameter space.}\label{Numan}

\subsection{Analysis strategy.} 

As has been already in the previous section a look-out for an additional vacuum at high scales 
requires one to study the evolution of the the various interaction strengths under under renormalization group equations. The 
values of the quartic couplings at the electroweak scale are, of course, connected with the masses and mixing angles in the scalar sector. A careful measurement of the signal strengths of the 
125 GeV at the LHC has revealed that the
resonance has couplings strikingly similar to the SM ones. These observations
have their ramifications on the 2HDM parameter space. Thus together with the requirement of 
having $m_h \simeq 125$ GeV, we also
arrange for $\b - \a \simeq \frac{\pi}{2}$, in order to comply with these results from
the LHC ($\b - \a = \frac{\pi}{2}$ is the well known \emph{alignment limit}\cite{PhysRevD.92.075004} in a 2HDM,
in which the couplings of $h$ to fermions and gauge bosons become exactly equal to the 
SM ones). We choose to describe a 2HDM parameter point in terms of the parameters
($\tan\beta, m_h, m_H, m_A, m_{H^+}, \a$) basis. There are additional constraints to satisfy
as outlined in the next few subsections,

\subsubsection{Perturbativity, unitarity and vacuum stability}
For the 2HDM to remain a perturbative quantum field theory at a given
energy scale, one must impose the conditions $\lvert \lambda_{i} \rvert
\leq 4\pi~(i=1,\ldots,7)$ and $ \lvert y_{i} \rvert \leq
\sqrt{4\pi}~(i=t,b,\tau)$ at that scale\footnote{The conditions are
  slightly different for the two types of couplings. The reason
  becomes clear if we note that the perturbative expansion
  parameter for $2 \rightarrow 2$ processes driven by the quartic
  couplings is $\lambda_i$. The corresponding parameter for
  Yukawa-driven scattering processes is $|y_i|^2$}. This translates
  into upper bounds on the running couplings at low as well as high scales.

A more sophisticated version of such bounds comes from the the requirement of partial wave unitarity
in longitudinal gauge boson scattering.
The 2$\rightarrow$2 amplitude matrix corresponding to scattering of the longitudinal components
of the gauge bosons can be mapped to a corresponding matrix for the scattering of the goldstone 
bosons\cite{Akeroyd:2000wc,Horejsi:2005da}. The theory respects unitarity if each eigenvalue of the aforementioned amplitude
matrix does not exceed 8$\pi$.
\besub
\bea
a_{\pm}&=& \frac32(\lambda_1+\lambda_2)\pm \sqrt{\frac94 (\lambda_1-\lambda_2)^2+(2\lambda_3+\lambda_4)^2},\\
b_{\pm}&=& \frac12(\lambda_1+\lambda_2)\pm \sqrt{\frac14 (\lambda_1-\lambda_2)^2+\lambda_4^2},\\
c_{\pm}&=& d_{\pm} = \frac12(\lambda_1+\lambda_2)\pm \sqrt{\frac14 (\lambda_1-\lambda_2)^2+\lambda_5^2},\\
e_1&=&(\lambda_3 +2\lambda_4 -3\lambda_5),\\
e_2&=&(\lambda_3 -\lambda_5),\\
f_1&=& f_2 = (\lambda_3 +\lambda_4),\\
f_{+}&=& (\lambda_3 +2\lambda_4 +3\lambda_5),\\
f_{-}&=& (\lambda_3 +\lambda_5).
\eea
\label{e:LQTeval}
\eesub

When the quartic part of the scalar
potential preserves CP and $\mathbb{Z}_2$ symmetries, the aforementioned
eigenvalues are discussed in
\cite{Kanemura:1993hm,Akeroyd:2000wc,Horejsi:2005da}.  

The condition to be discussed next is that of vacuum stability.
For the scalar potential of a theory to be stable, it must be bounded
from below in all possible directions. This condition is threatened if the quartic part
of the scalar potential, which is responsible for its behaviour at large
field values, turns negative. Avoiding such a possibility till any given scale
ensures stability of the vacuum up to that scale. Vacuum stability in context of
a 2HDM has been discussed in detail in \cite{Barroso:2013awa}

Demanding high-scale positivity of the 2HDM potential along various directions in the field space
leads to the following conditions on the scalar potential.
\cite{Branco:2011iw,Ferreira:2004yd}, 
\besub 
\bea
\label{e:vsc1}
\rm{vsc1}&:&~~~\lambda_{1} > 0 \\
\label{e:vsc2}
\rm{vsc2}&:&~~~\lambda_{2} > 0 \\
\label{e:vsc3}
\rm{vsc3}&:&~~~\lambda_{3} + \sqrt{\lambda_{1} \lambda_{2}} > 0 \\
\label{e:vsc4}
\rm{vsc4}&:&~~~\lambda_{3} + \lambda_{4} - |\lambda_{5}| + \sqrt{\lambda_{1} \lambda_{2}} > 0
\label{e:vsc5}
\eea
\label{eq:vsc}
\eesub
When a second vacuum arises, we go for a test of metastability of the electroweak vacuum using
vsc2 alone. The reason for this has already been discussed at the end of section 3.

\subsubsection{Oblique parameters and flavor constraints.}
A 2HDM contributes to the electroweak precision obsevables through the participation of the
additional scalars in loops. For example, the oblique $S$, $T$ and $U$ parameters receive
contributions $\Delta S$, $\Delta T$ and $\Delta U$ repectively from the 2HDM. The most 
constraining amongst these is  $\Delta T$ which reads,

\besub
\bea
 \Delta T &=& F(m_{H^+}^2,m_{H}^2)+F(m_{H^+}^2,m_{A}^2)+c_{\beta - \alpha}^2 F(m_{H^+}^2,m_{h}^2)+s_{\beta - \alpha}^2 F(m_{H^+}^2,m_{H}^2)-F(m_{H}^2,m_{A}^2)\nonumber \\
 & &
-c_{\beta - \alpha}^2 F(m_{h}^2,m_{A}^2)-s_{\beta - \alpha}^2 F(m_{H}^2,m_{A}^2)+3c_{\beta - \alpha}^2(F(m_{Z}^2,m_{H}^2)-F(m_{W}^2,m_{H}^2))\nonumber \\
 & &
-3c_{\beta - \alpha}^2(F(m_{Z}^2,m_{h}^2)-F(m_{W}^2,m_{h}^2)) 
\eea
\eesub

Where,

\be
F \left( m_{1}^2, m_{2}^2 \right) \equiv \left\{ \begin{array}{lcl}
\displaystyle{
\frac{m_{1}^2 + m_{2}^2}{2} - \frac{m_{1}^2 m_{2}^2}{m_{1}^2 - m_{2}^2}\, \ln{\frac{m_{1}^2}{m_{2}^2}}
}
& ; & m_{1}^2 \neq m_{2}^2,
\\*[3mm]
0 & ; & m_{1}^2 = m_{2}^2.
\end{array} \right.
\label{funcF}
\ee

We have filtered the model points through the constraint $\Delta T = 0.05 \pm 0.12$ following \cite{Baak:2014ora}. On the restrictions coming from the flavor physics side, measurement of the $b \rightarrow s \gamma$ leads to $m_{H^+} \geq 315$ GeV in case 
of the Type-II 2HDM\cite{Mahmoudi:2009zx}. In case of Type-I, there is no such lower bound. The constraint $m_{H^+} \geq 80$ GeV originating from direct searches however still persists.

\subsubsection{$h\rightarrow\gamma\gamma$ decay width}

For a near-alignment case, the Higgs signal strengths to 
The partial decay width of the SM-like Higgs to a pair of photons in this case has the expression\cite{Gunion:1989we},
\begin{eqnarray}
 \Gamma^{2HDM} (h\to \gamma\gamma) = \frac{\alpha^2g^2}{2^{10}\pi^3}
 \frac{m_h^3}{M_W^2} \Big|sin(\b-\a) F_W + \Big(\frac{\cos\a}{\sin\b}\Big)\frac{4}{3}F_t  + \kappa F_{{H^{+}}} \Big|^2
 \,, 
\end{eqnarray}

The functions ${F}_W$, ${F}_t$ and ${F}_{i+}$ encapsulate the effects of a W-boson, a t-quark
and a charged scalar running in the loop and shall be defined as,
\begin{subequations}
\begin{eqnarray}
 F_W &=& 2+3\tau_W+3\tau_W(2-\tau_W)f(\tau_W) \,,  \\
 F_t &=& -2\tau_t \big[1+(1-\tau_t)f(\tau_t)\big] \,,  \\
 F_{H^+} &=& -\tau_{i+} \big[ 1-\tau_{i+}f(\tau_{i+}) \big]\,.
\end{eqnarray}
\end{subequations}
\begin{eqnarray}
f(\tau) &=&
\left[\sin^{-1}\left(\sqrt{1/\tau}\right)\right]^2 \,.\\ \rm with,~ 
\tau &=& \frac{4 m^2_{a}}{m^2_h}
\label{f}
\end{eqnarray}

Here, $a$ = $t$, $W$ and $H^{+}$.

We assume $h$ is dominanly produced through gluon fusion. 
In such a case, the signal strength for the diphoton final state is approximately
given by,

\bea
\mu_{\gamma \gamma} &=& \frac{\Gamma^{2HDM} (h\to \gamma\gamma)}{\Gamma^{SM} (h\to \gamma\gamma)}
\eea

In order to respect the 2$\sigma$ bound on $\mu_{\gamma \gamma}$, from a combined
measurement of ATLAS and CMS, we discard model points that violate
$\mu_{\gamma \gamma} \in $ [1.04,~1.37] \cite{Aad:2014eha}

\section{Results and discussions.}\label{Results}  

Model points that successfully negotiate all the aforesaid constraints are allowed to evolve under RG,
till some scale $\L$ (say). $\L$ is essentially identified as the scale where perturbative
unitarity is destroyed, and can be interpreted as the scale up to which no physics over and above
the extended Higgs sector is required. If there is an additional, lower vacuum before $\L$, the time scale for tunneling from the EW vacua to the new one must therefore be larger than the age of the 
universe.
 It is intuitively expected that higher is $\L$, tighter becomes
the parameter space that is allowed at the electroweak scale. This is indeed confirmed by
the findings reported in REF. Of course the points leading to a metastable EW vacuum are 
identified through a detailed scan of the parameter space. However, the fate of a particular
model point at high scales is sensitive to the value of the top quark mass taken. With this in 
view, we propose the benchmarks listed in Table~\ref{BP}.

%\begin{table}[h]
%\centering
%\begin{tabular}{|c c c c c c c c|}
%\hline
%Benchmark  & tan$\beta$ & $m_{H}$(GeV) & $m_{A}$(GeV) & $m_{H^+}$(GeV) & $m_{12}$(GeV)  & Perturbative till \\ \hline \hline
%
%BP1 & 1.779 & 354.425 & 380.075 & 341.220 & 222.578 & $\sim10^{7}$ GeV \\ \hline
%BP2 & 2.501 & 489.800 & 506.900 & 486.570 & 286.043 & $\sim10^{11}$ GeV \\ \hline
%BP3 & 7.278 & 320.780 & 297.660 & 324.300 & 117.726 & $\sim10^{16}$ GeV \\ \hline
%BP4 & 8.282 & 500.844 & 500.590 & 500.446 & 172.791 & $\sim10^{19}$ GeV \\ \hline
%BP5 & 6.896 & 501.768 & 499.784 & 500.664 & 189.166 & $\sim10^{19}$ GeV \\ \hline
%BP6 & 10.94 & 1499.940 & 1500.000 & 1498.840 & 451.481 & $\sim10^{19}$ GeV \\ \hline
%\end{tabular}
%\caption{Benchmark points chosen to illustrate the behaviour under RGE. $\L$ denotes the maximum
%extrapolation scale up to which  perturbativity remains intact.}
%\label{BP}
%\end{table}

\begin{table}[h]
\centering
\begin{tabular}{|c c c c c c c c|}
\hline
Benchmark  & tan$\beta$ & $m_{H}$(GeV) & $m_{A}$(GeV) & $m_{H^+}$(GeV) & $m_{12}$(GeV)  & Perturbative till \\ \hline \hline

BP1 & 1.78 & 354 & 380 & 341 & 222 & $\sim10^{7}$ GeV \\ \hline
BP2 & 2.50 & 489 & 506 & 486 & 286 & $\sim10^{11}$ GeV \\ \hline
BP3 & 7.28 & 320 & 297 & 324 & 117 & $\sim10^{16}$ GeV \\ \hline
BP4 & 8.28 & 500 & 500 & 500 & 172 & $\sim10^{19}$ GeV \\ \hline
BP5 & 6.90 & 501 & 499 & 500 & 189 & $\sim10^{19}$ GeV \\ \hline
BP6 & 10.94 & 1499 & 1500 & 1498 & 451 & $\sim10^{19}$ GeV \\ \hline
\end{tabular}
\caption{Benchmark points chosen to illustrate the behaviour under RGE. $\L$ denotes the maximum
extrapolation scale up to which  perturbativity remains intact.}
\label{BP}
\end{table}

\begin{figure}[!htbp]
\begin{center}
\includegraphics[scale=0.45]{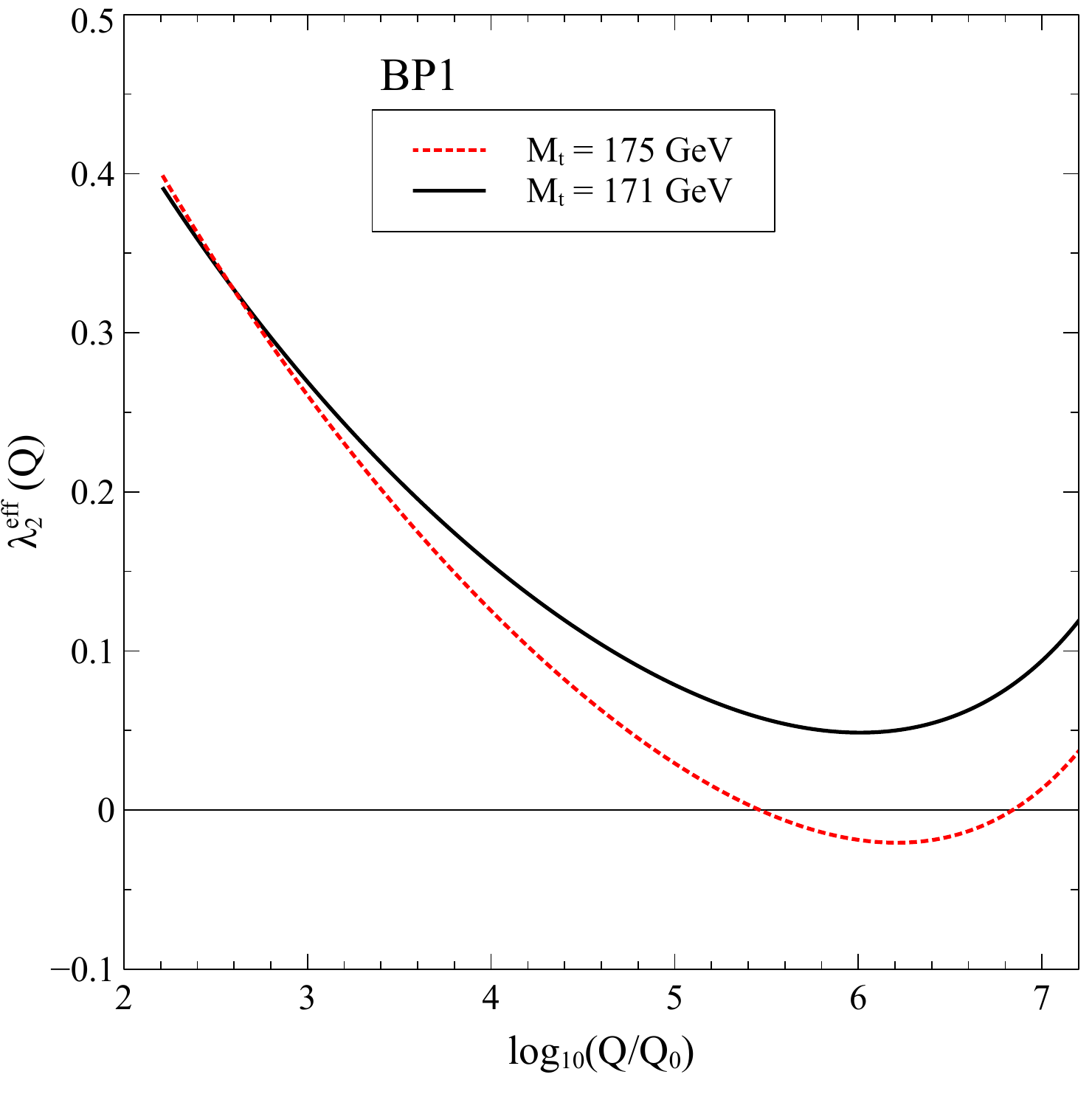}~~~~~~~~
\includegraphics[scale=0.45]{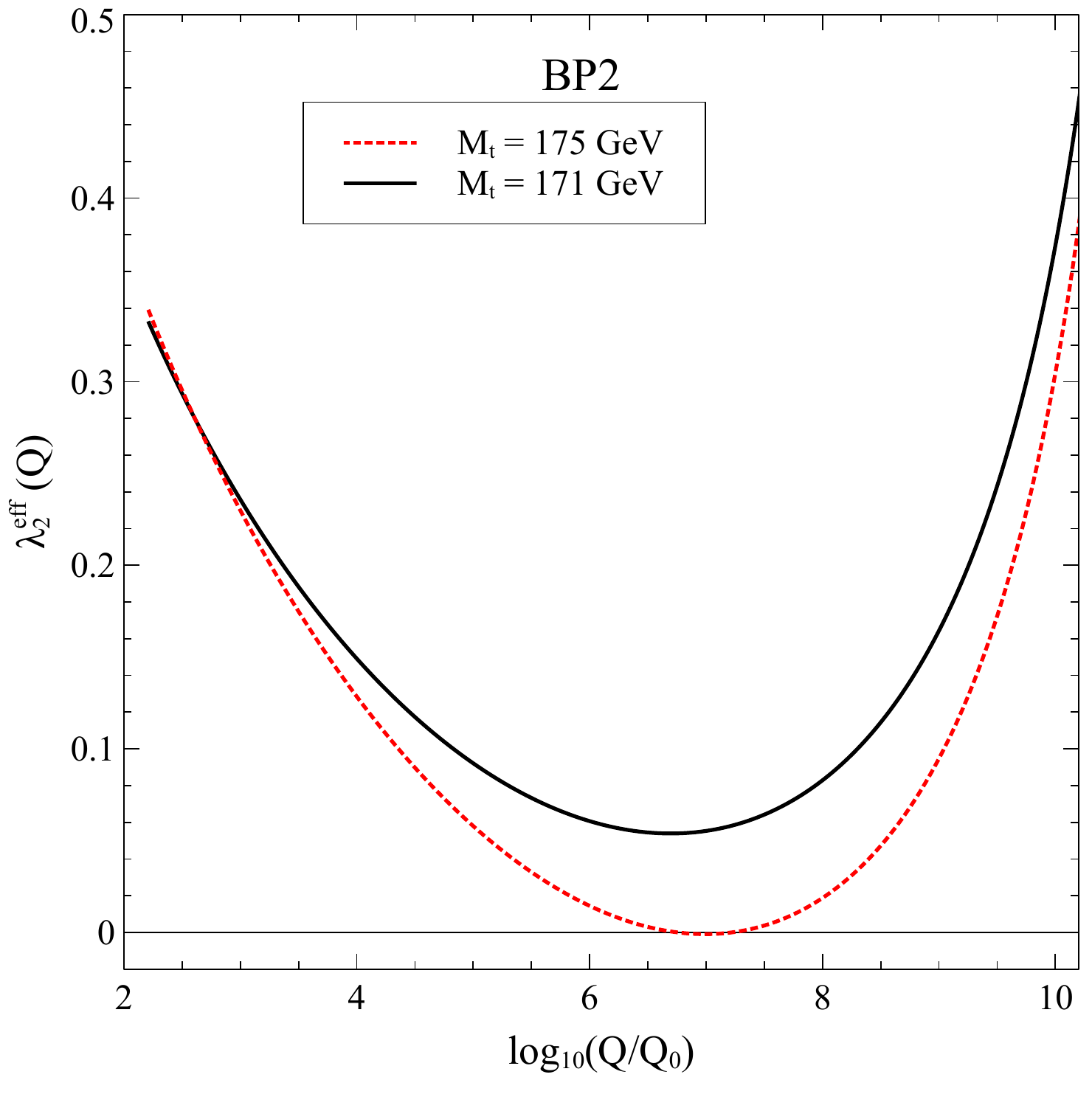}
\includegraphics[scale=0.45]{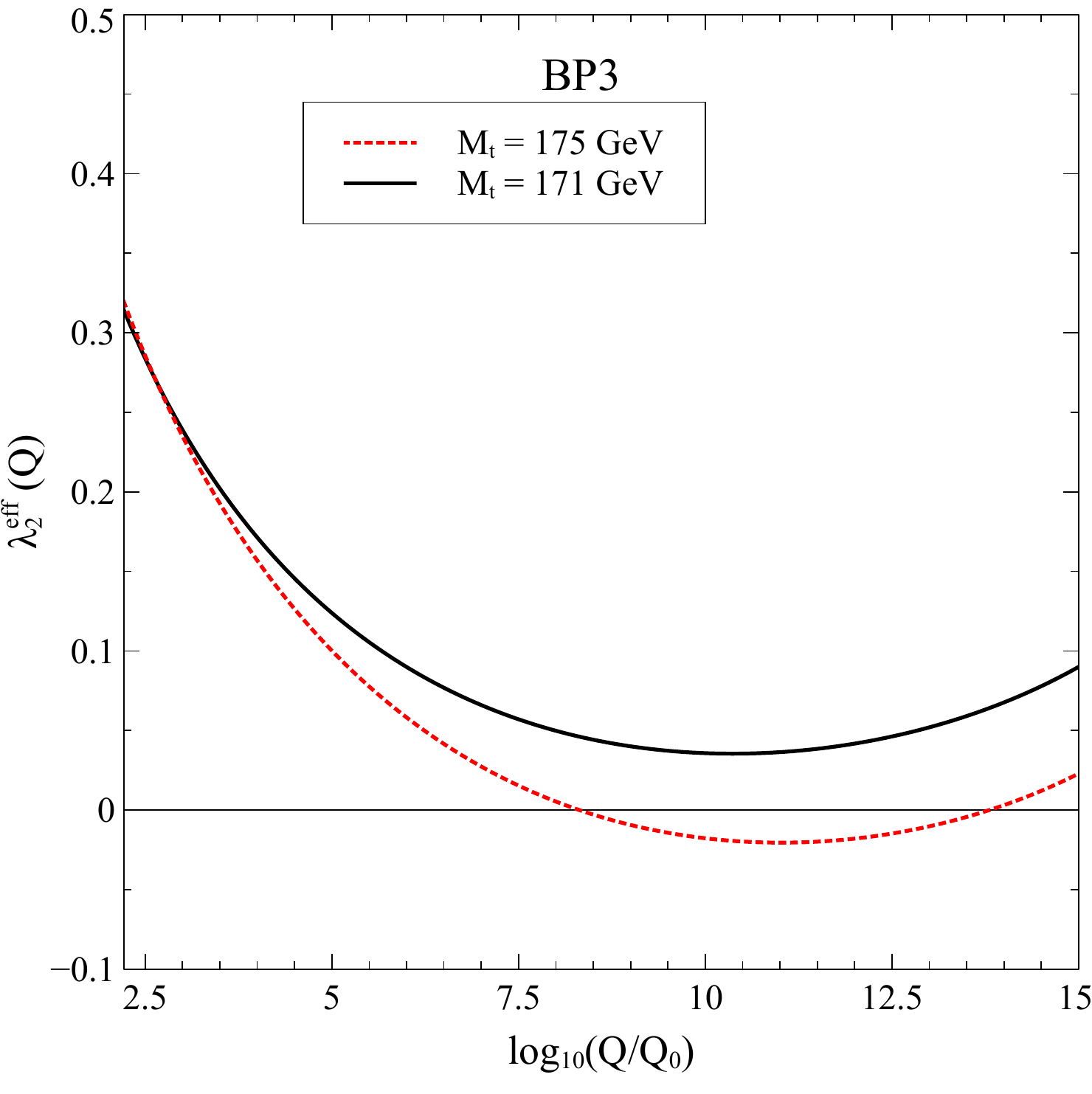}~~~~~~~~
\includegraphics[scale=0.45]{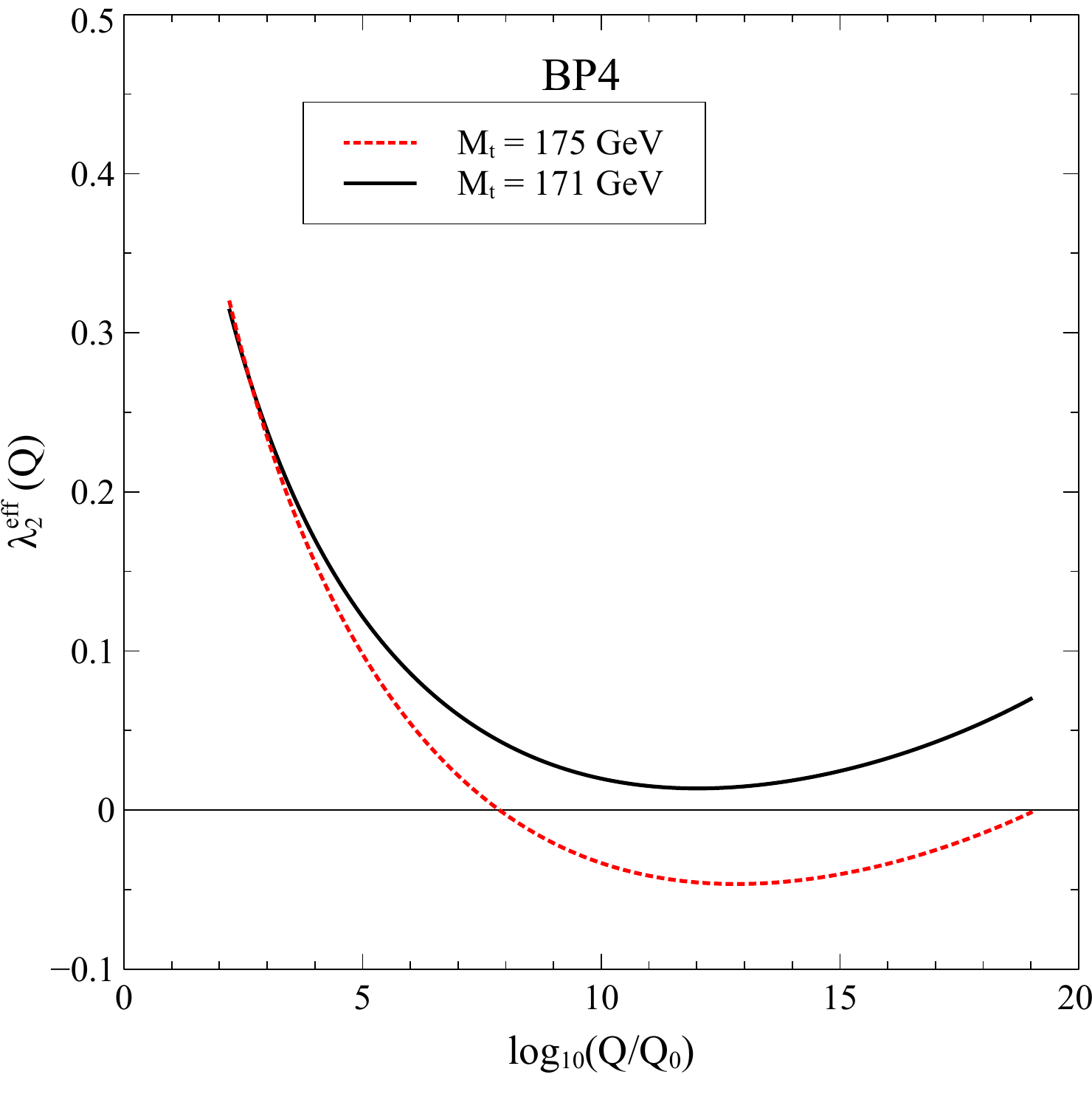}
\includegraphics[scale=0.45]{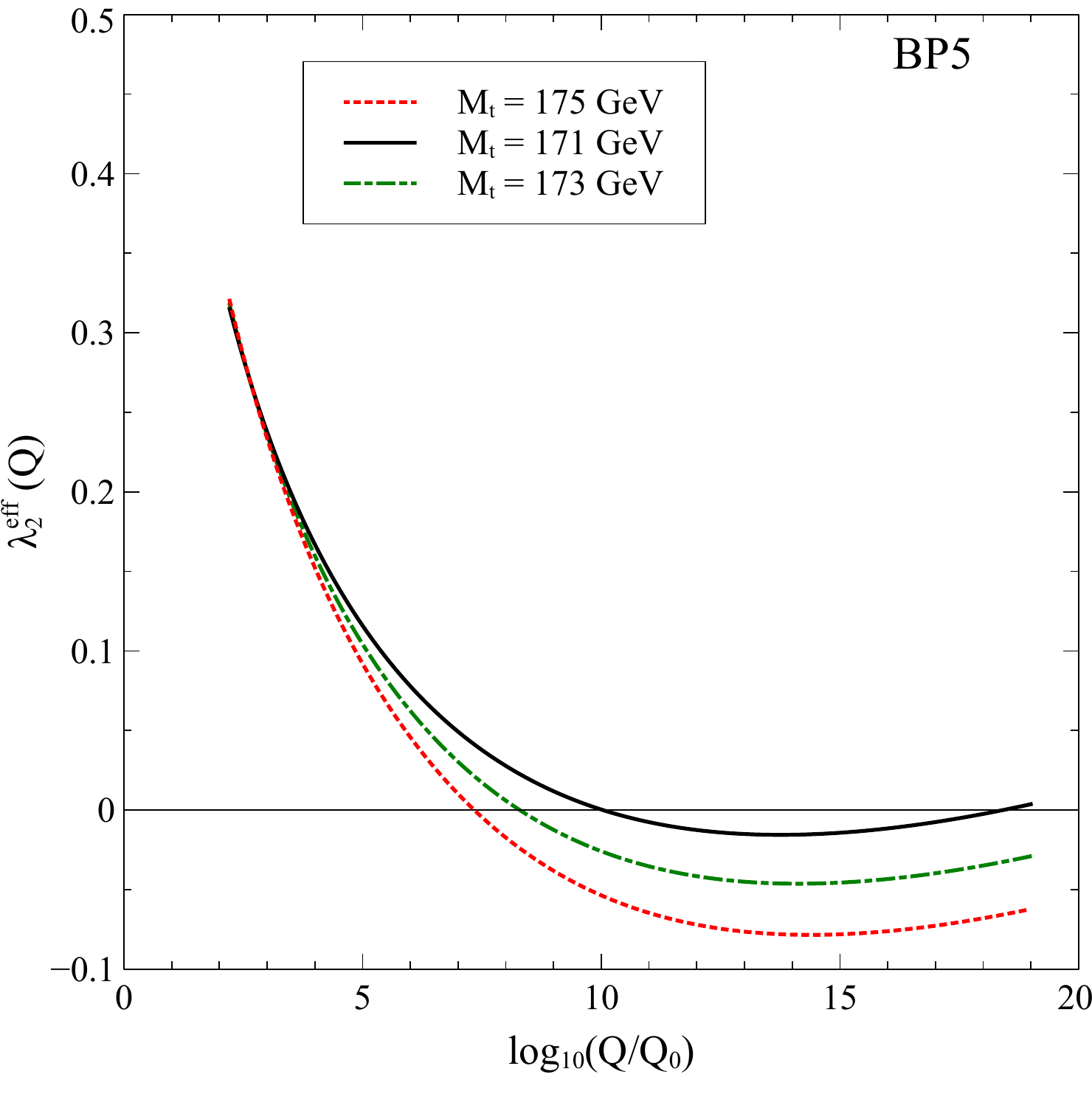}~~~~~~~~
\includegraphics[scale=0.45]{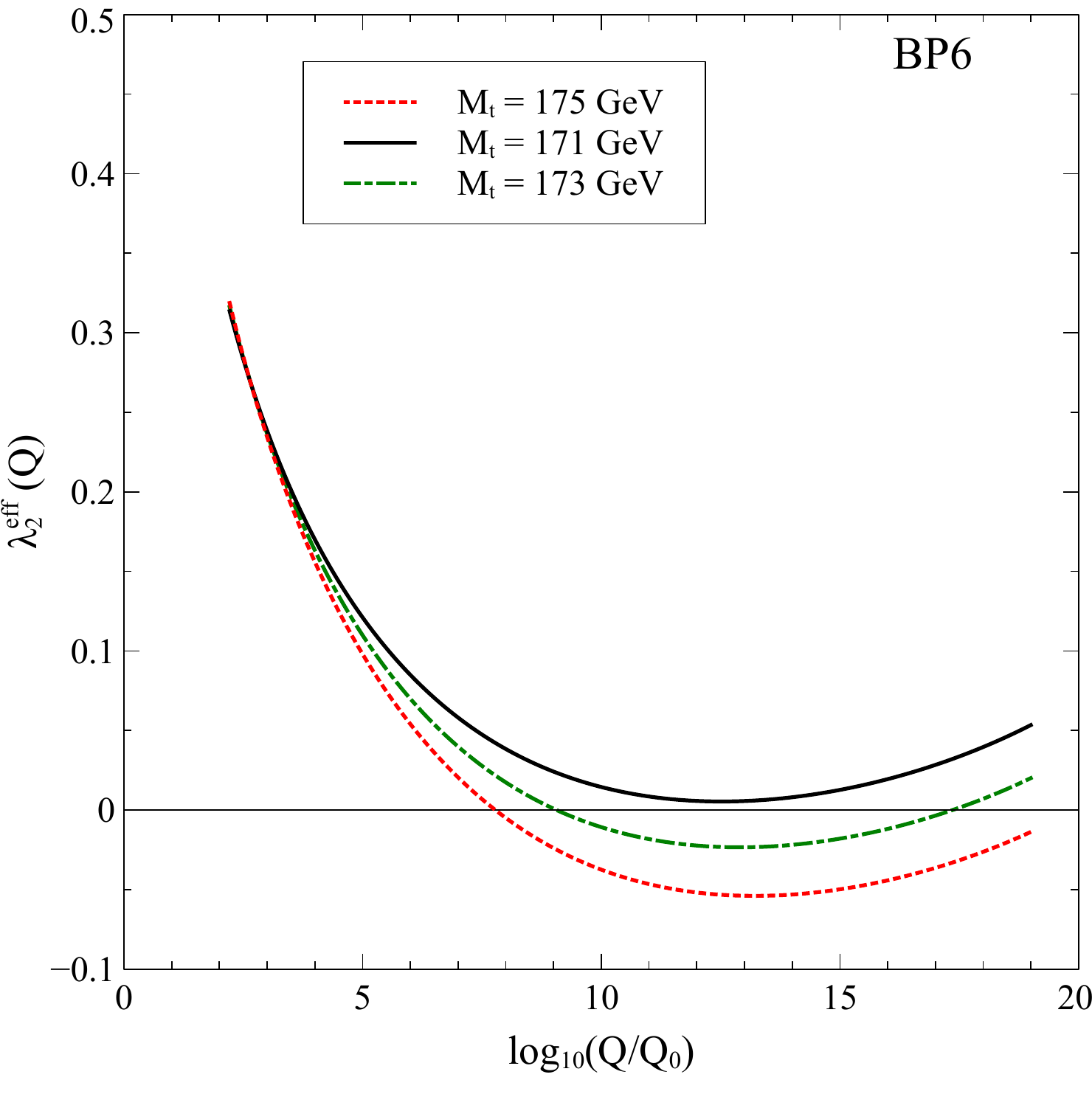}
\caption{RG evolution of $\l_2^{eff}$ for the benchmarks listed in Table~\ref{BP}, for more than
one value of $M_t$. 
The color coding is explained in the legends. }
\label{f:running}
\end{center}
\end{figure}

In each case, we plot the evolution of  $\l_2^{eff}$ in Fig~\ref{f:running}
The chosen benchmarks differ in their perturbative behaviour at high scales, although
all of them have the common feature that the EW vacuum turns metastable, or even unstable for $M_t$ = 175 GeV.
In other words, the Type-II 2HDM may turn non-perturbative beyond a scale $\L$, even though 
a vacuum deeper than the EW one might be encountered someplace intermediate between the electroweak scale and $\L$. Besides, although it is worth identifying
those parameter points that keep the EW vacuum metastable all the way till the GUT or Planck scales,
we also include for completeness in the benchmarks, two points where a 2HDM loses its perturbativity at a much lower scale.
For instance in BP1, at $M_t$ = 175 GeV, $\l_2^{eff}$ turns negative and, $\frac{d \l_2^{eff}}{dt}$ = 0 occurs around $6.2 \times 10^{6}$ GeV (The scale at which the tunneling probability gets maximized).
Using Eqn, one obtains $\l_2^{eff} < -0.05$ in this case. An inspection of Fig~\ref{f:running} thus indicates that
this particular benchmark leads to metastability. The same parameter point offers absolute stability
for $M_t$ = 171 GeV though. BP2 describes the same qualitative feature as BP1, albeit it remains
perturbative till $10^{11}$ GeV.   

\begin{figure}[!htbp]
\begin{center}
\includegraphics[scale=0.45]{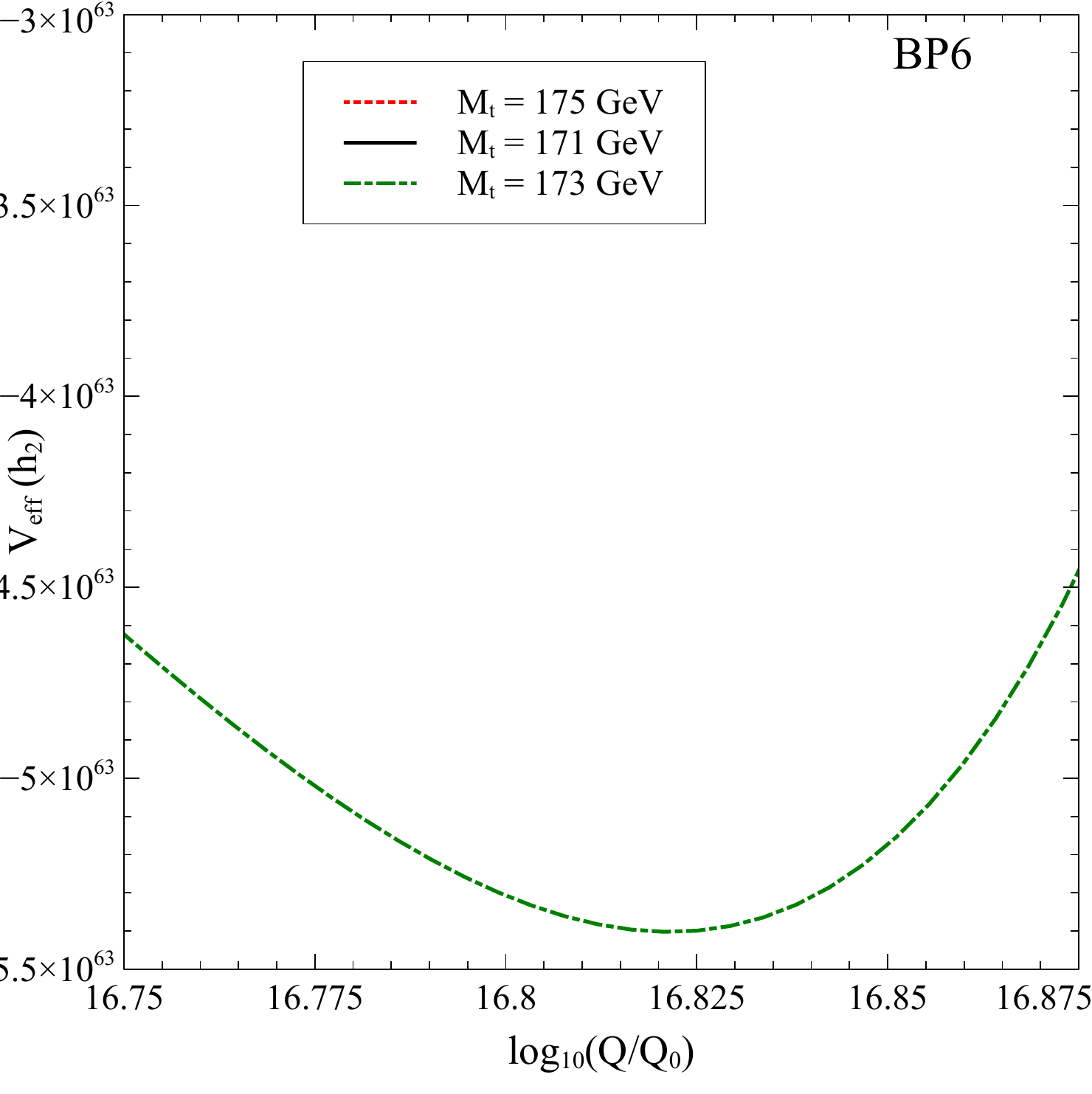}~~~~~~~~
\caption{Behaviour of $V_{eff}(h_2)$ in BP6 for $M_t$ = 173 GeV.}
\label{f:pot}
\end{center}
\end{figure}

BP3 is a more conservative benchmark in the sense that, it keeps the 2HDM perturbative
till the GUT scale and also prevents an \emph{unstable} EW vacuum even in the worst case scenario 
with $M_t$ = 175 GeV. We remind the reader that the strength of the top quark Yukawa coupling depends not
only on the pole mass, but also on tan$\beta$. This becomes crucial in deciding the fate of the EW
vacuum at high scales. For instance, BP5 experiences a higher t-quark negative pull compared to BP4 owing to a lower value of tan$\beta$ in BP5, even though the quartic couplings at the input scale
are at the same ball-park for the two cases.   
BP6 is a fine-tuned parameter point that is perturbative till
the Planck scale, and for which the EW vacuum is stable, metastable or unstable for $M_t$ = 171 GeV, 
173 GeV and 175 GeV respectively. For the sake of completeness, we display
the behaviour $V_{eff}(h_2)$ for $M_t$ = 173 GeV case in Fig~\ref{f:pot}.

Model points are randomly sampled in the following specified ranges,
\bea
\tan\beta \in [0.1, 20.0] \nonumber \\
m_H, m_A, m_{H^+}, m_{12} \in [0,1200 ~GeV]
\eea

\begin{figure}[!htbp]
\begin{center}
\includegraphics[scale=0.40]{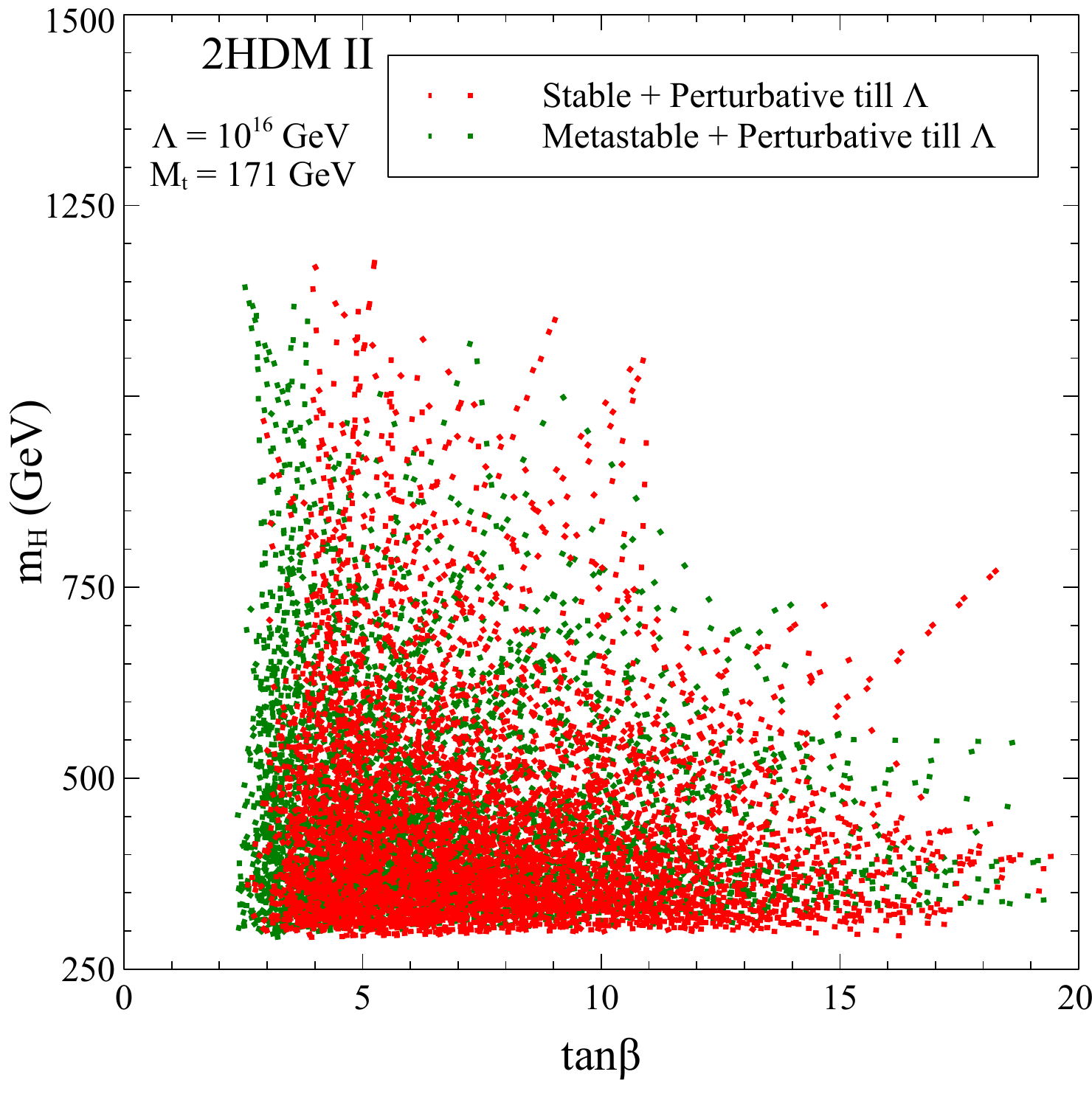}~~~~~~
\includegraphics[scale=0.40]{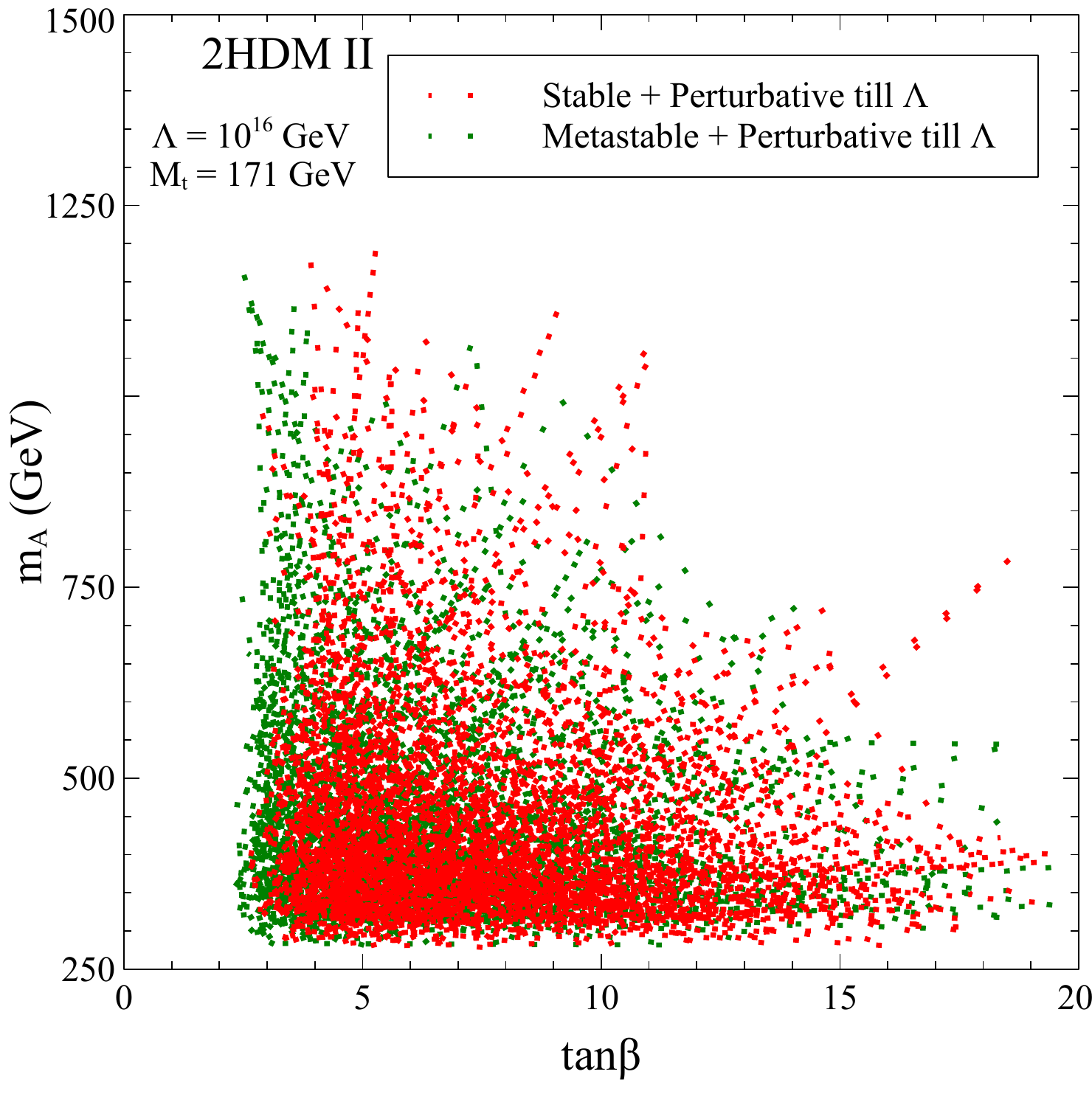}~~~~~~
\includegraphics[scale=0.40]{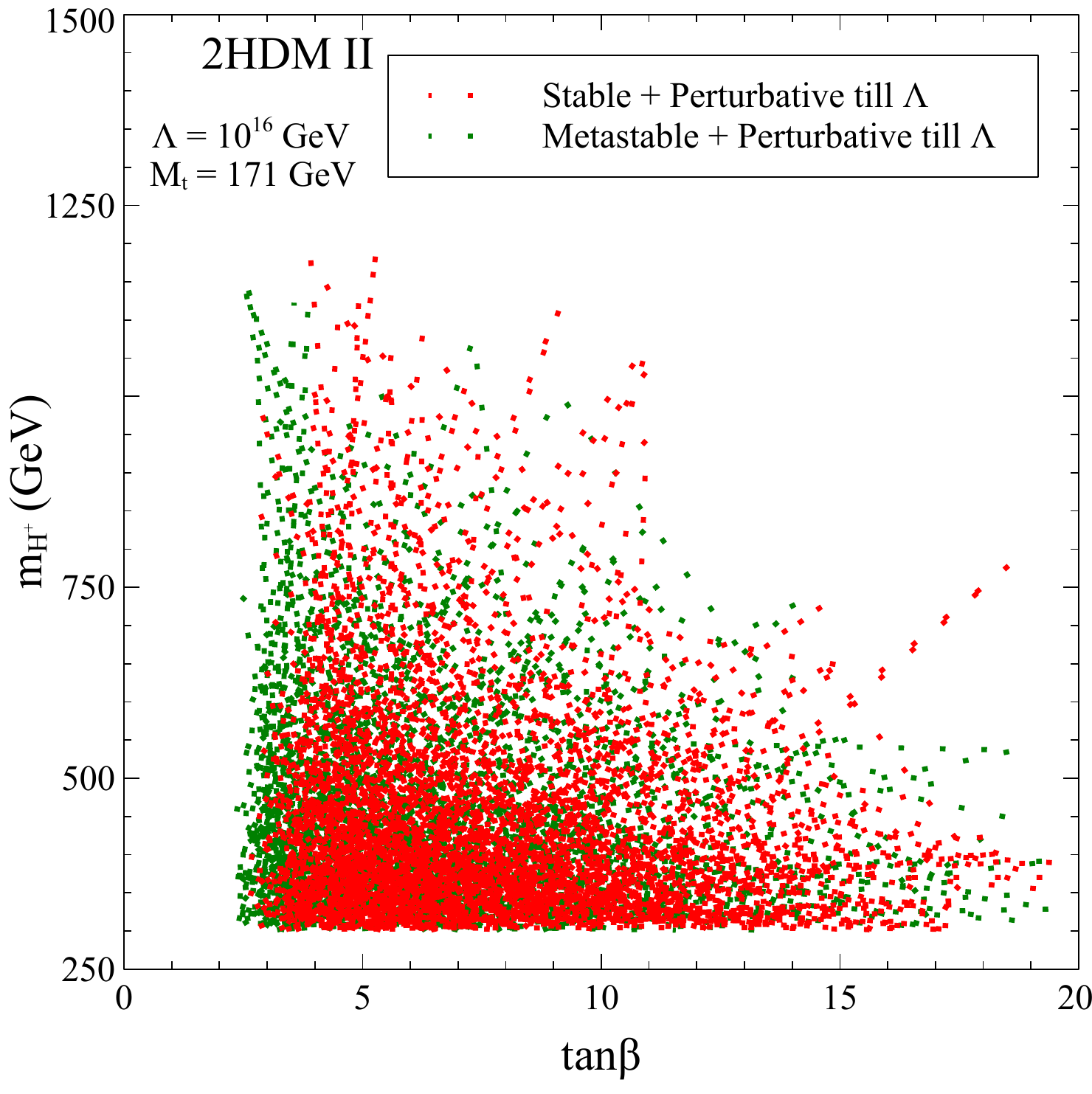}
\includegraphics[scale=0.40]{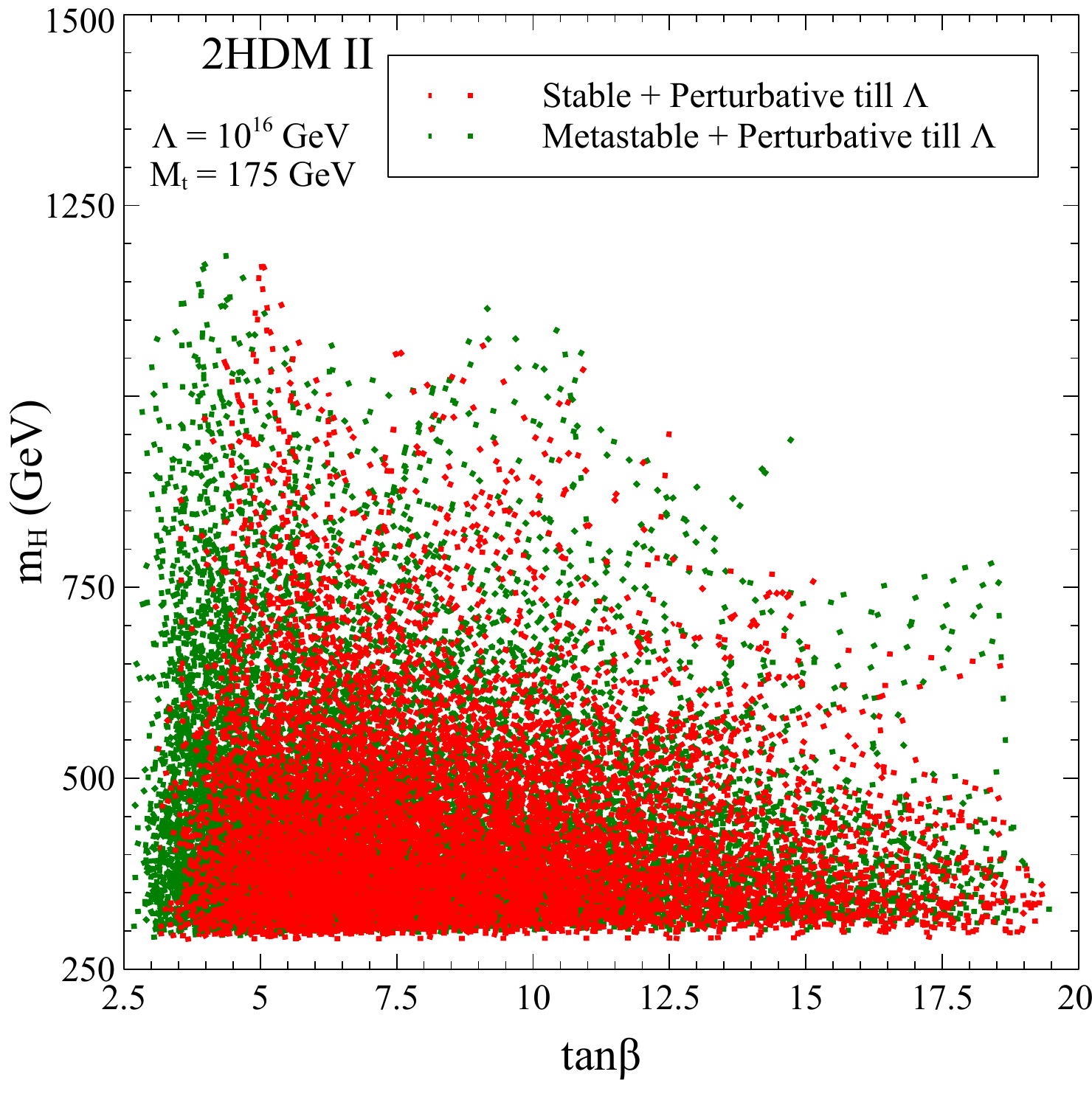}~~~~~~
\includegraphics[scale=0.40]{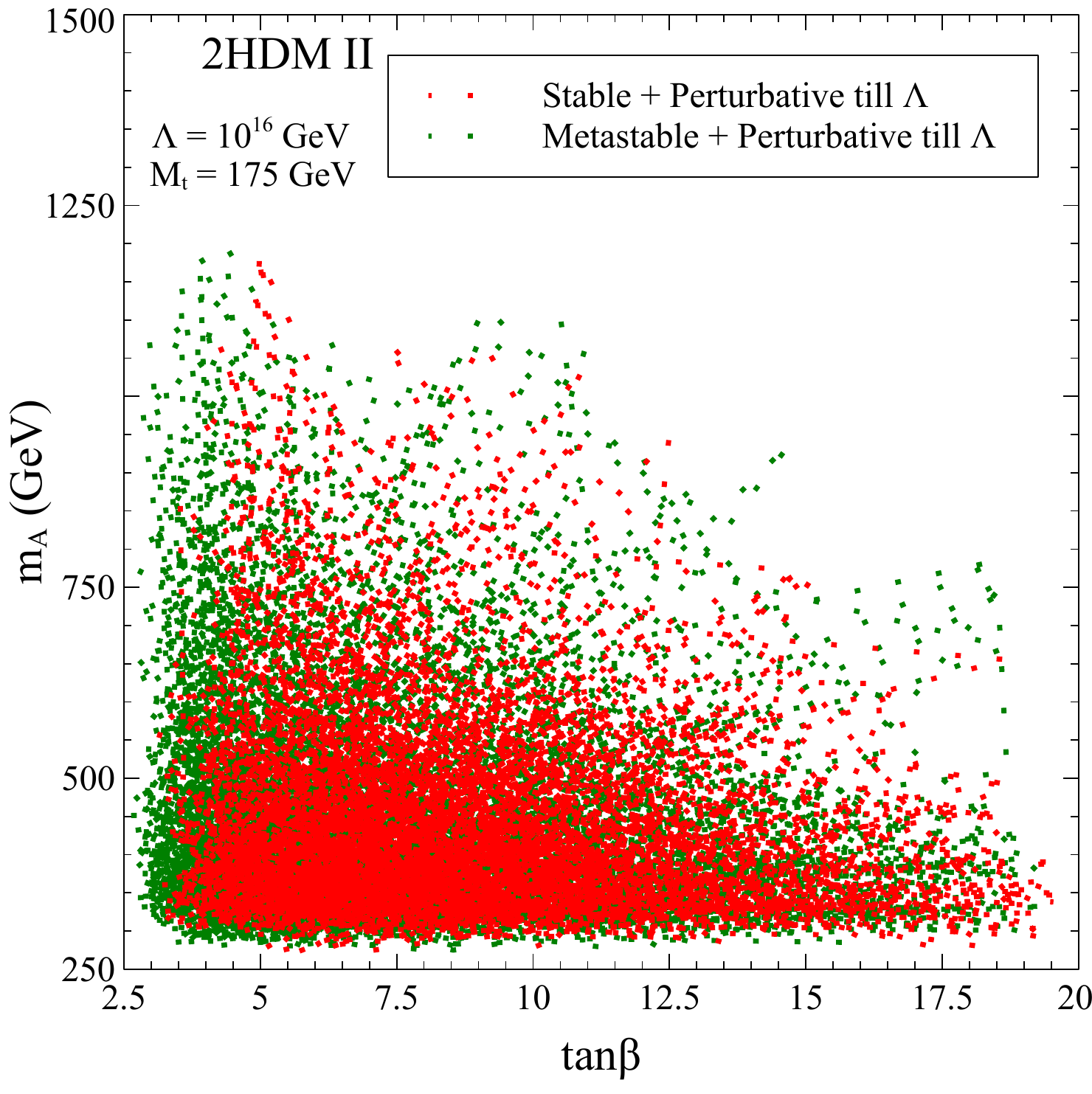}~~~~~~
\includegraphics[scale=0.40]{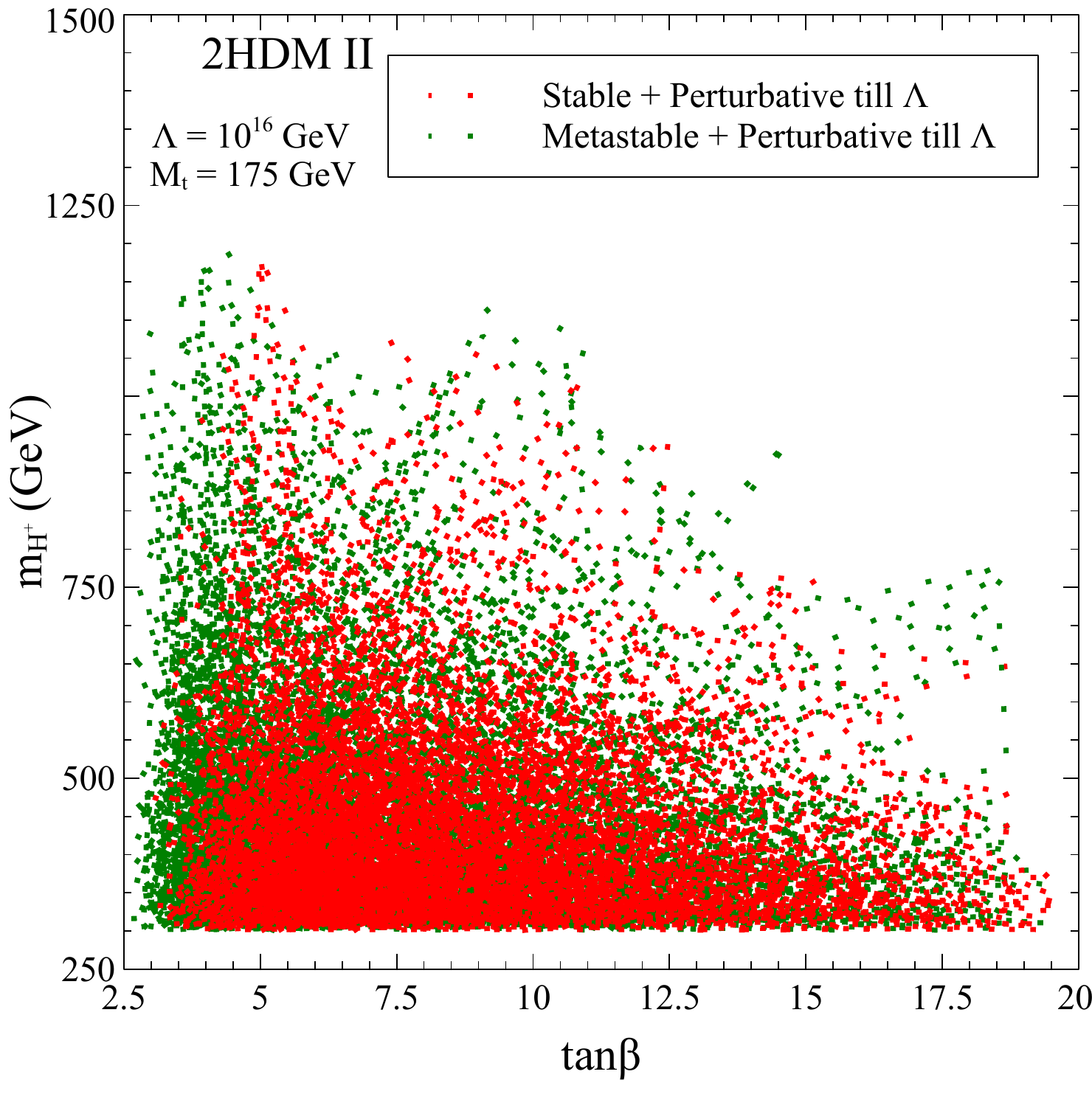}

\caption{Distribution of models perturbative till $10^{16}$ GeV that lead to stable, or 
metastable EW vacuum. The upper (lower) plots correspond to $M_t$ = 171 (175) GeV. The color
coding is explained in the legends. 2HDM II refers to a Type II 2HDM.}
\label{f:tb-m_E16}
\end{center}
\end{figure}

\begin{figure}[!htbp]
\begin{center}
\includegraphics[scale=0.40]{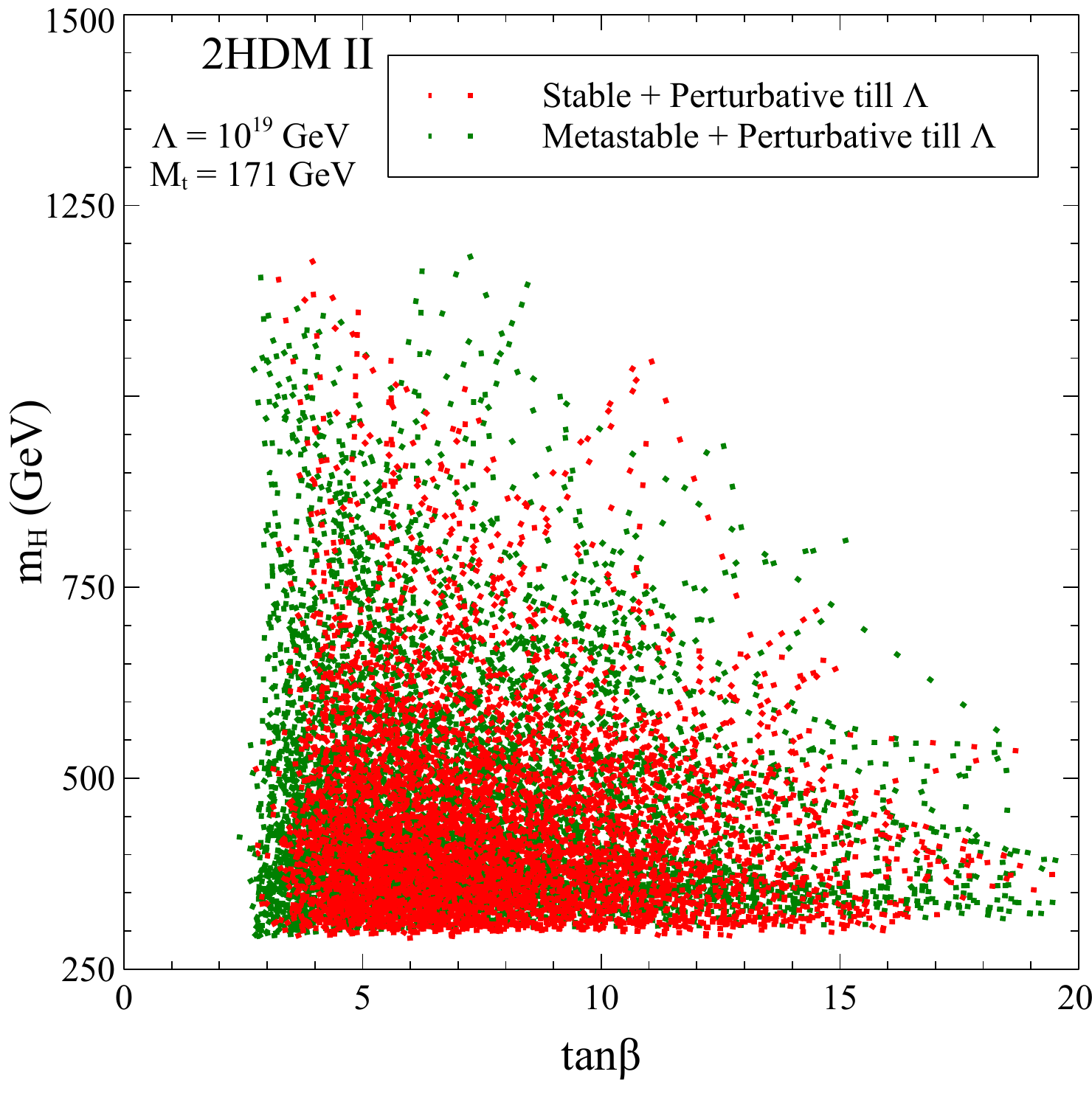}~~~~~~
\includegraphics[scale=0.40]{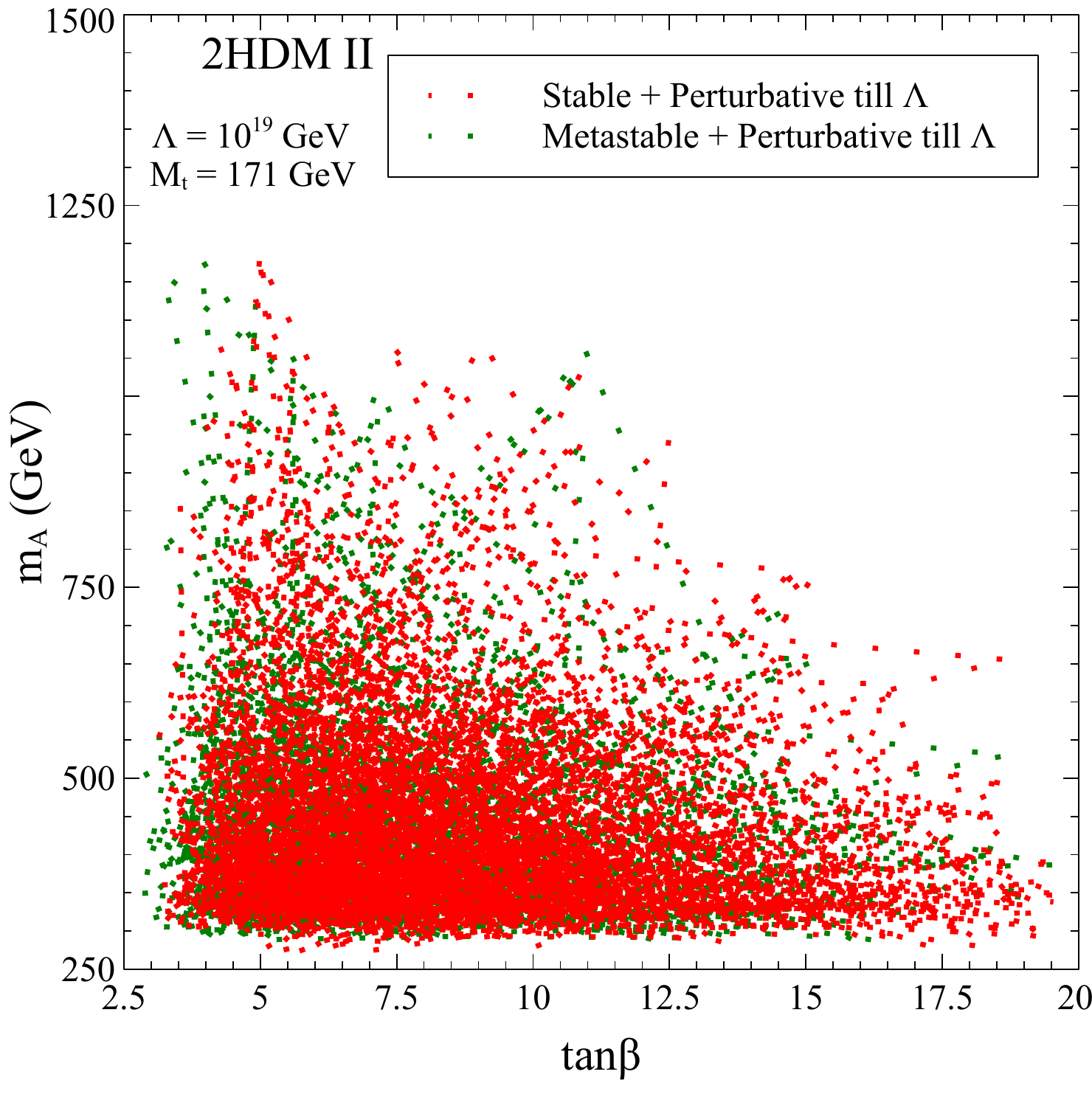}~~~~~~
\includegraphics[scale=0.40]{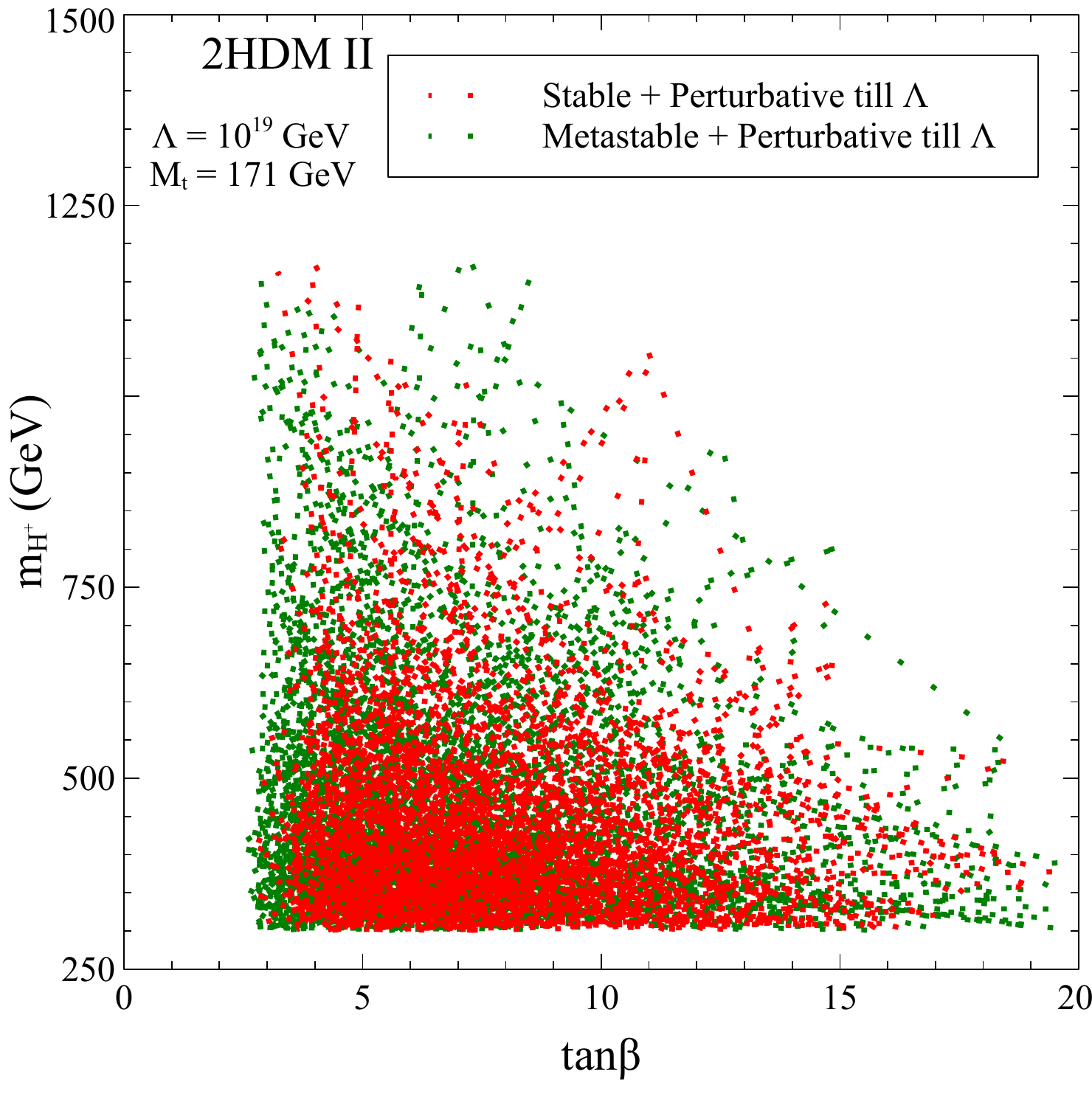}
\includegraphics[scale=0.40]{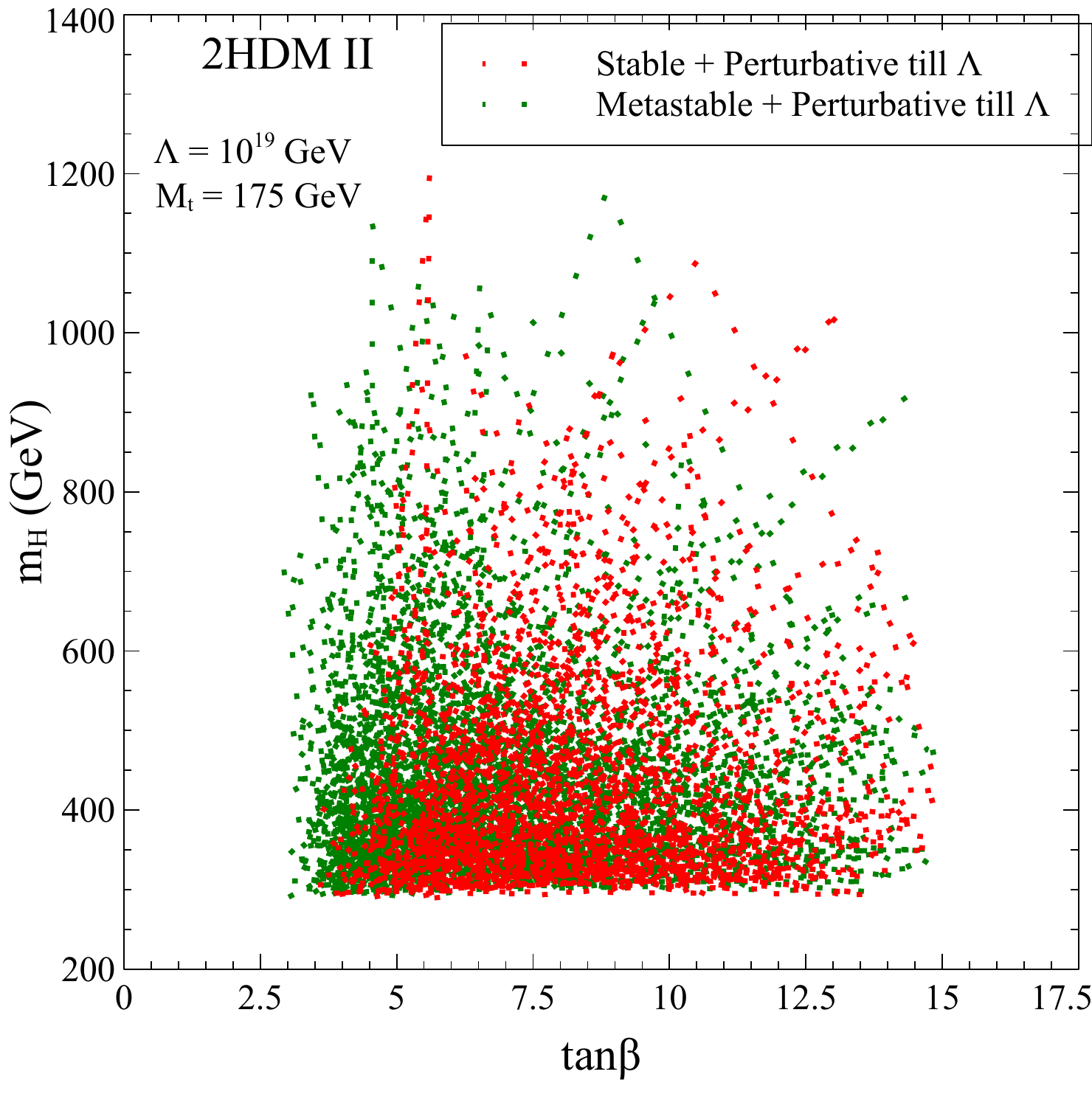}~~~~~~
\includegraphics[scale=0.40]{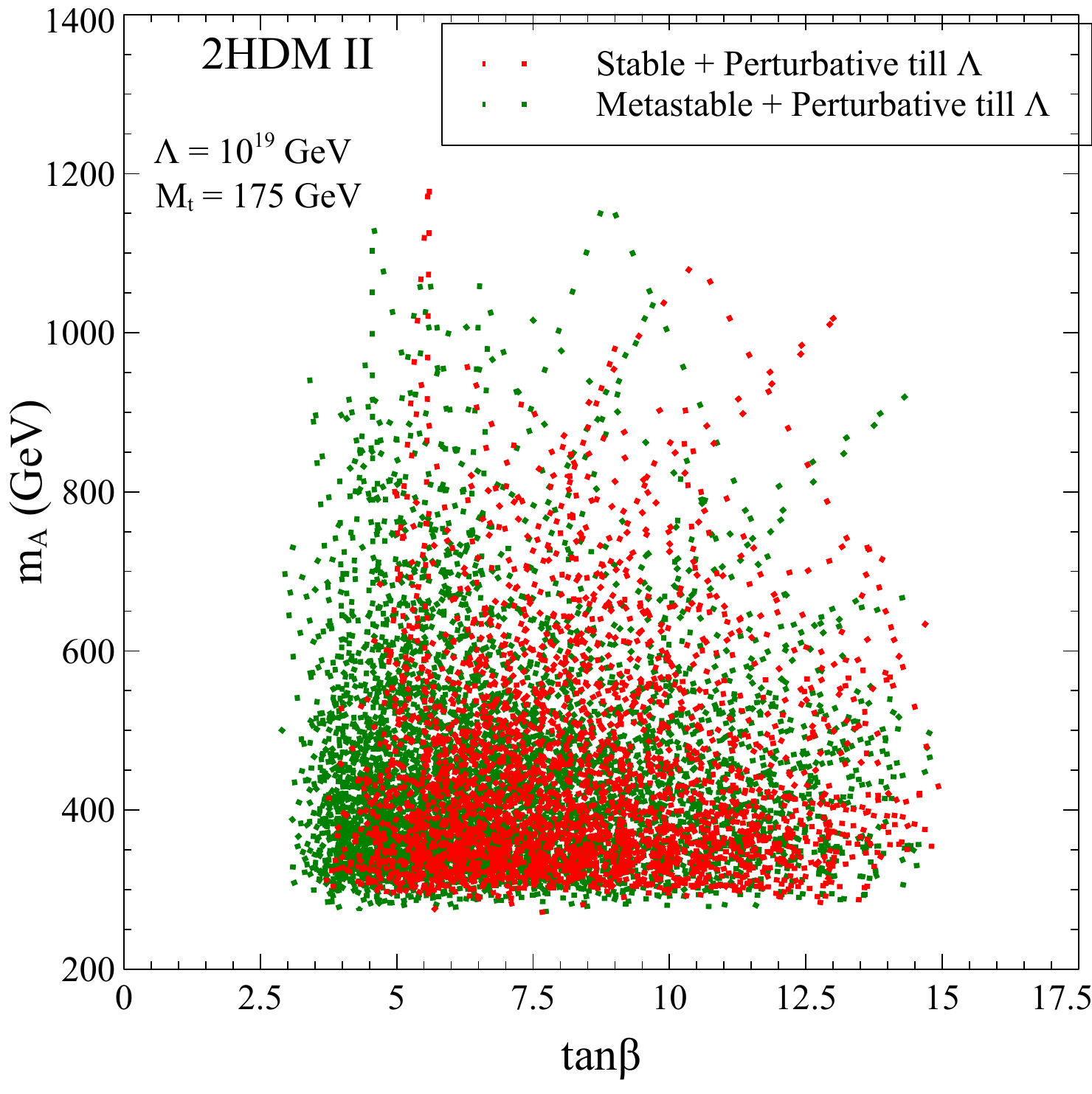}~~~~~~
\includegraphics[scale=0.40]{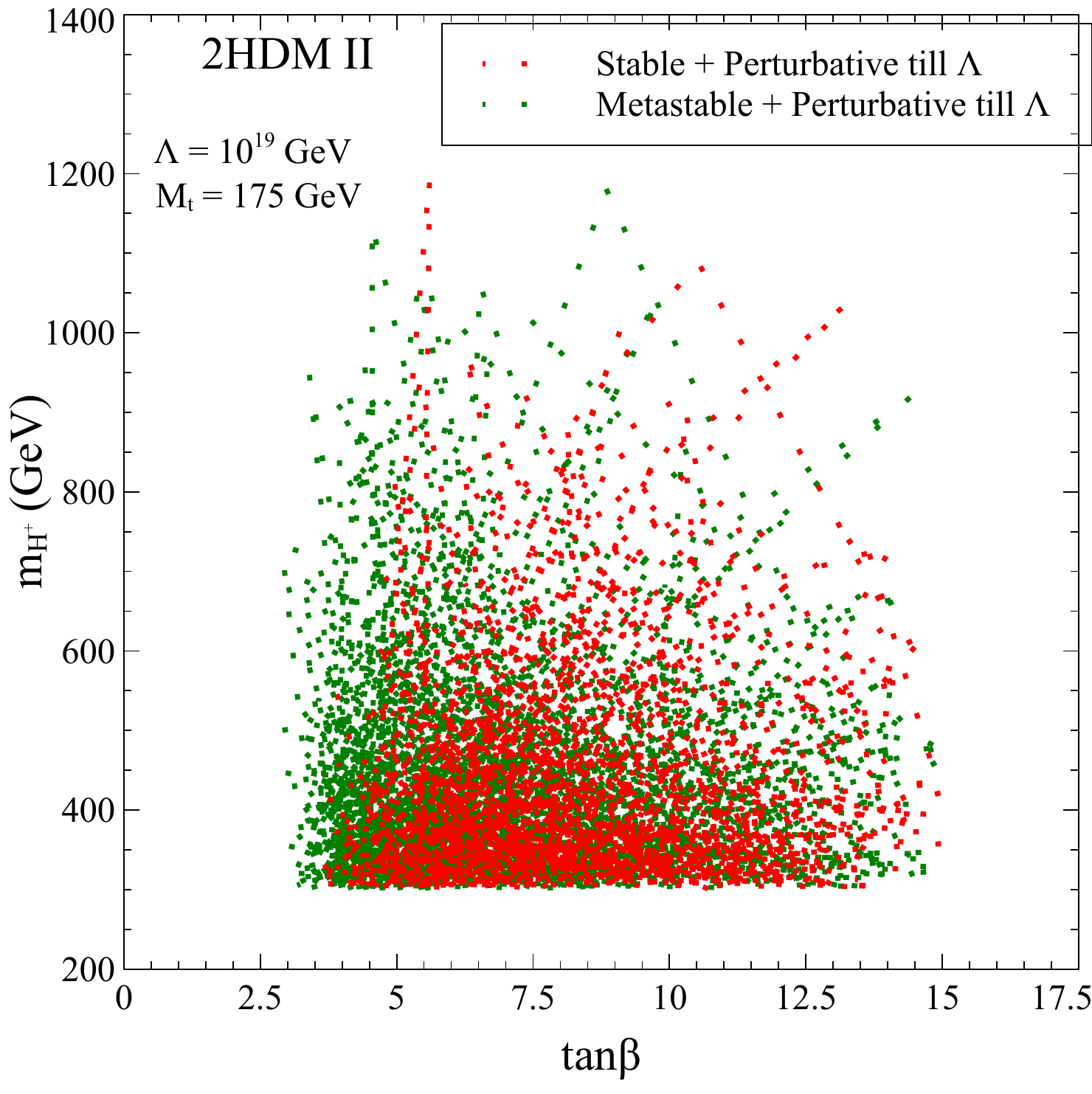}

\caption{Distribution of models perturbative till $10^{19}$ GeV that lead to stable, or 
metastable EW vacuum. The upper (lower) plots correspond to $M_t$ = 171 (175) GeV. The color
coding is explained in the legends. 2HDM II refers to a Type II 2HDM.}
\label{f:tb-m_E19}
\end{center}
\end{figure}

A condition forbidding the loss of perturbativity/unitarity at scale $\L$ is imposed throughout the scan.
The following broad features emerge from Fig~\ref{f:tb-m_E16} and Fig.~\ref{f:tb-m_E19}
The results are shown for Fig. 3 onwards 
two representative values of $\L$, namely $10^{16}$ GeV (Fig.3) and 
$\L$ = $10^{19}$ GeV (Fig.4).

(i) Perturbativity puts stringent constraints on the splitting amongst the masses. In fact,
for $\L$ = $10^{19}$ GeV the masses are near-degenerate. This effect can be attributed to the
fact that for a large mass a splitting, the couplings are already large at the input scale,
leading to a blow-up soon after. An individual mass does nor get bounded from the above however.
Thus, a perturbative theory till high scale automatically respects the T-parameter constraint.\\

(ii) A smaller tan$\beta$ for the same $M_t$ implies an enhanced 
fermionic contribution to the evolution of $\l_2$, and hence it favors 
a metastable vacuum over an absolutely stable one. Consequently, tan$\beta$ is bounded
from below in order to prevent tunelling to the lower vacuum. Moreover, one would
apprehend that the bound by obtained by demanding absolute stability of the EW vacuum
to be the stronger than the one obtained when one allows for a metastable scenario.
For instance, for $M_t$ = 171 GeV, the lower bounds read $\simeq$ 2.1 and $\simeq$ 2.5
for the two cases.\\
(iii) The lower bound on tan$\beta$ of course depends on the choice of $M_t$. 
For instance the parameter point parametrised in terms of the masses and tan$\beta$
indeed shall have different evolution trajectories for two different values of $M_t$.
This is reflected in the upper and lower plots of Fig~\ref{f:tb-m_E16}, where one
witnesses a tighter lower bound, for both the "stable" as well as "metastable" models.
Of course, in this case too, absolute stability yields a stronger bound than metastability.
For this value of $M_t$, any model with tan$\beta < $ 2.6 yields a tunneling lifetime lower
than the age of the universe.\\
(iv) Although the lower bound on tan$\beta$ should also depend on the $\L$ chosen, it hardly
changes with respect to the $10^{16}$ GeV value for $\L$ = $10^{19}$ GeV. Only the number of
allowed points shrinks to some extent, other essential features are unchanged.

In the plane of tan$\beta$ vs masses, it is expected that a particular parameter point responsible for
a metastable EW vacuum can always be found in the vicinity of a point that leads to absolute 
stability, tan$\beta$ $\geq$ 3.0. This gets confirmed by an inspection of Fig.~\ref{f:tb-m_E19}. 
 This can be understood from the fact that any enhanced fermionic contribution due to a higher 
tan$\beta$ can always be cancelled by an appropriately increased bosonic contribution through a slight tweak in the masses . Of course, one also has to
keep the couplings perturbative in doing so. Such a "fine-tuned" existence of a metastable EW vacuum
is not a surprise and is always expected in the case of an extended Higgs sector, such as the 2HDM. 

\begin{figure}[!htbp]
\begin{center}
\includegraphics[scale=0.40]{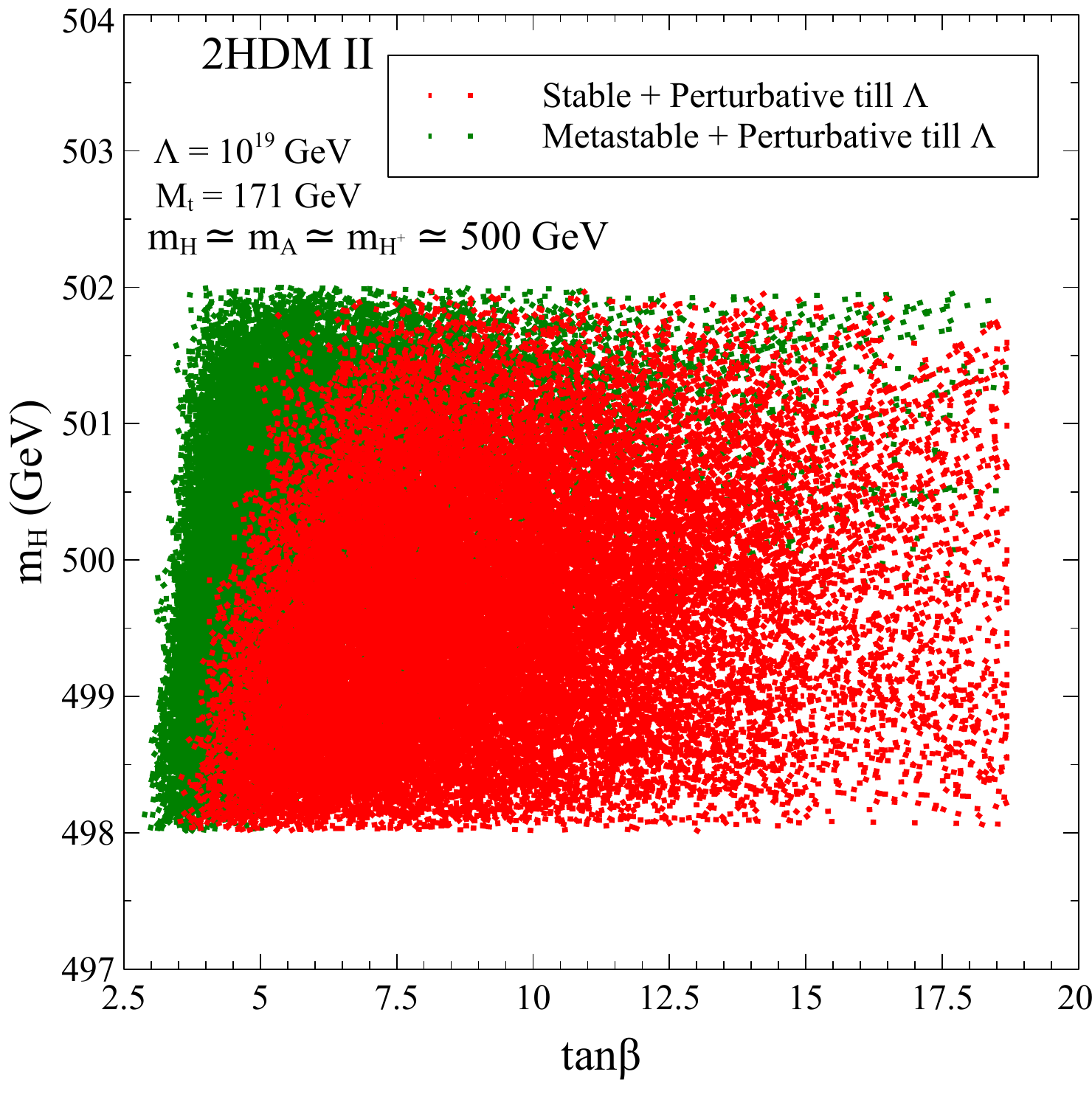}~~~~~~
\includegraphics[scale=0.40]{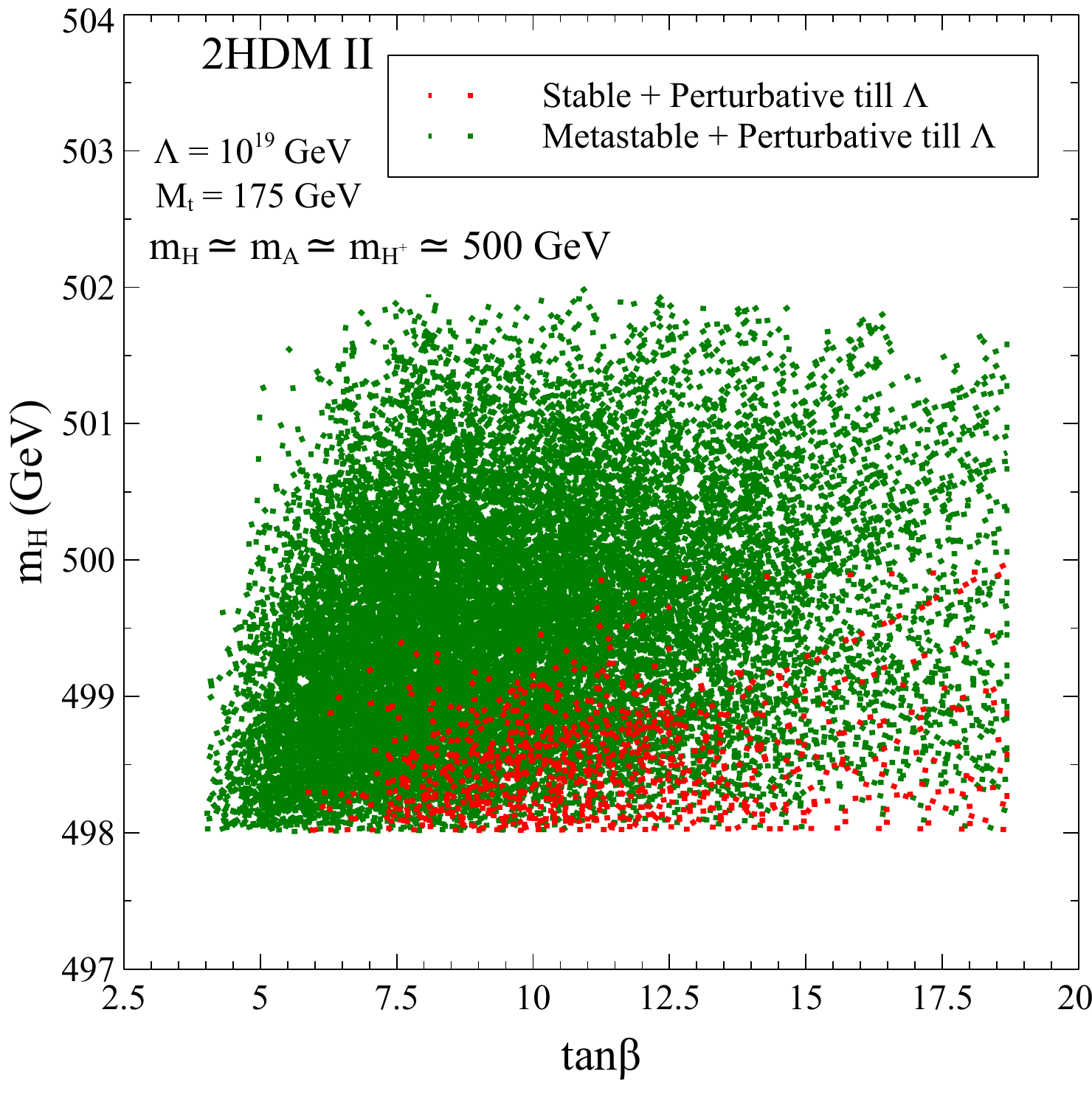}
\includegraphics[scale=0.40]{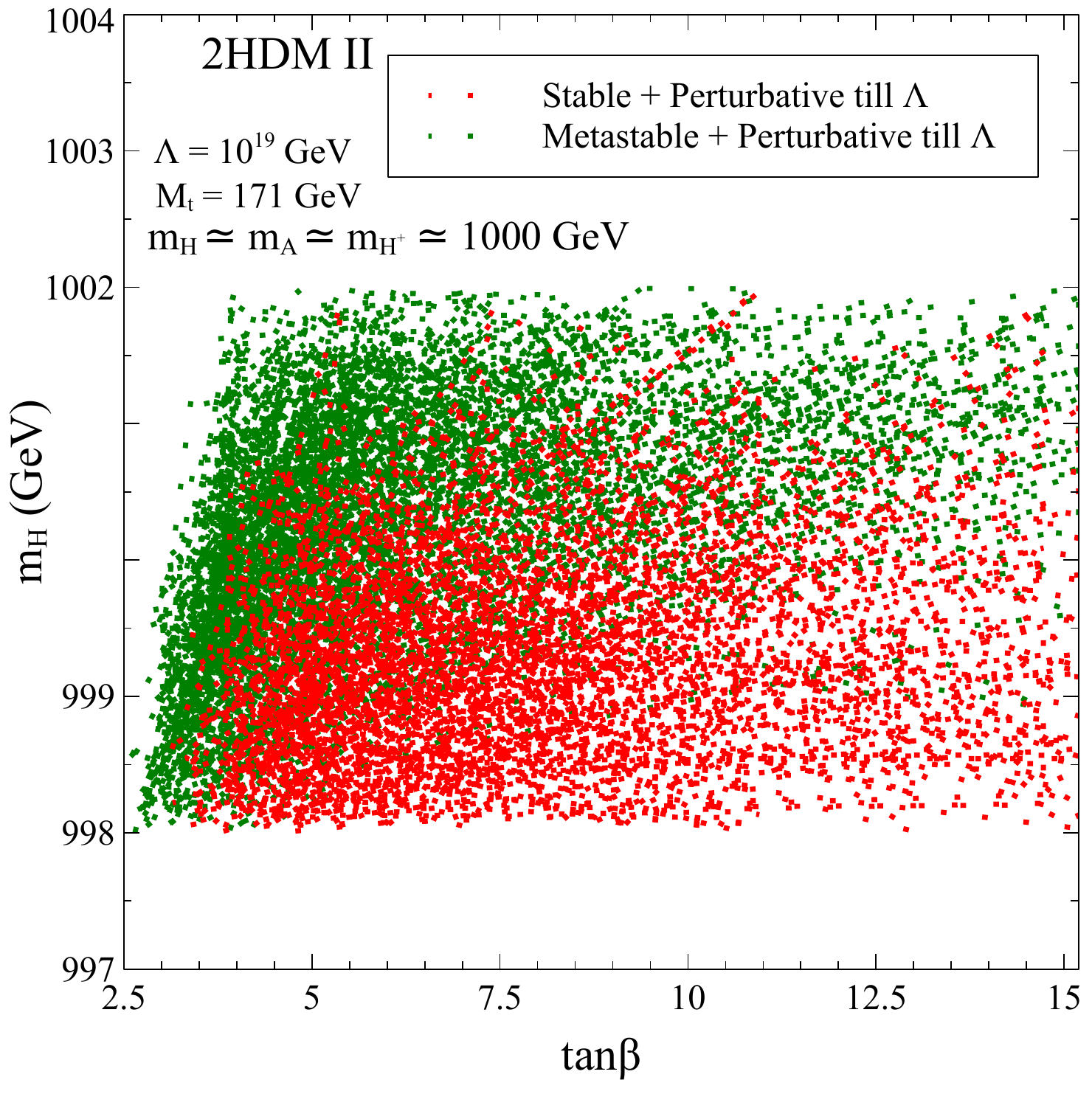}~~~~~~
\includegraphics[scale=0.40]{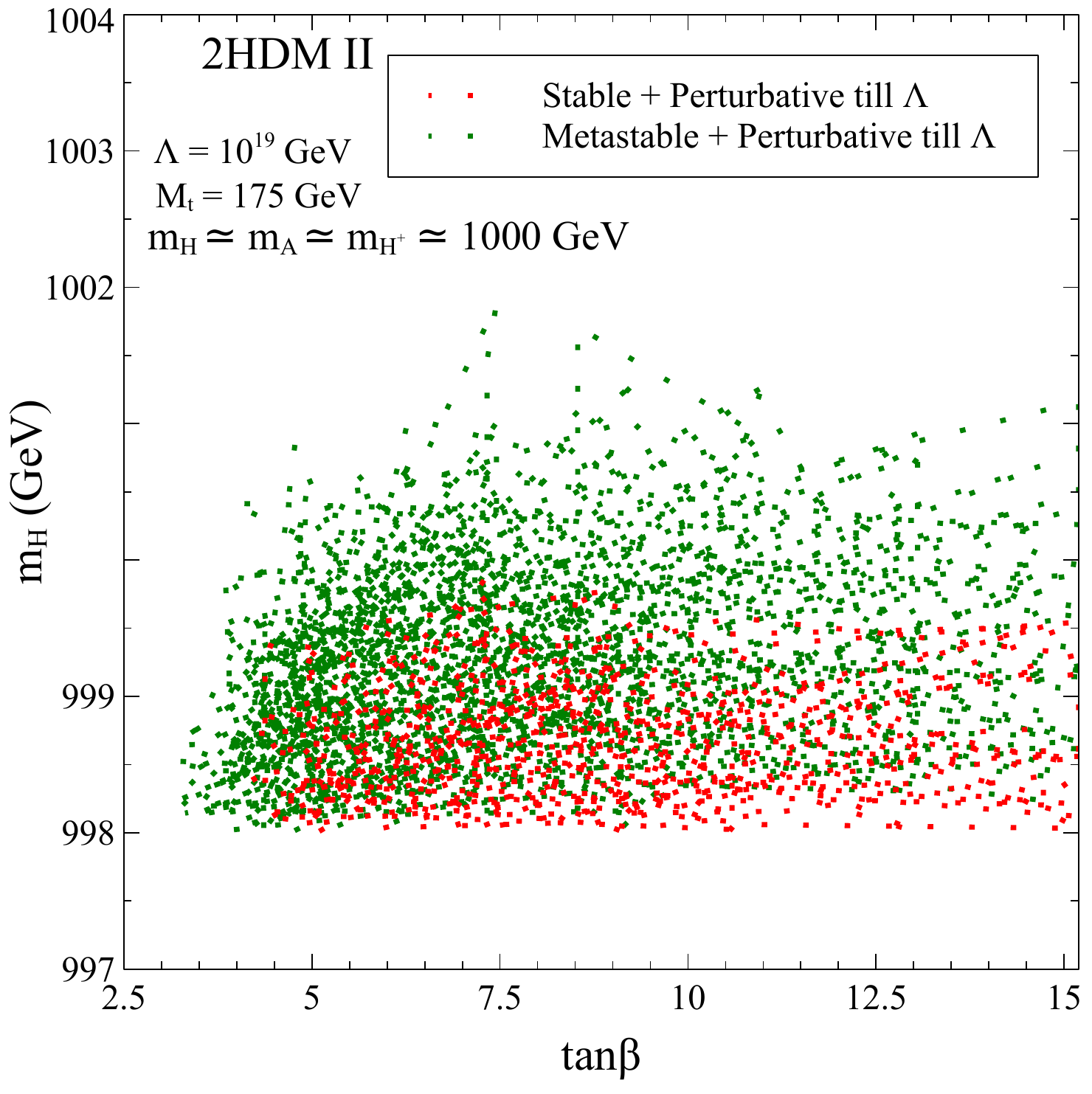}

\caption{ Distribution of models perturbative till $10^{19}$ GeV that lead to stable, or 
metastable EW vacuum. The upper (lower) plots correspond to $M_t$ = 171 (175) GeV. The color
coding is explained in the legends. 2HDM II refers to a Type II 2HDM.}
\label{f:mfixed}
\end{center}
\end{figure}

We take another approach where different scalar masses are fixed within specific narrow ranges,
 and allow tan$\beta$ to vary. This approach
turns useful in demarcating the "stable" region from the "metastable". We thus propose two
central values of 500 GeV and 1000 GeV and allow only a 2 GeV split about that.
Fig.~\ref{f:mfixed} presents the results for this choice. 

For masses around 500 GeV and $M_t$ = 171 GeV, the metastable points mostly cluster in the low tan$\beta$ region. They
get largely disfavored at larger tan$\beta$. Since the bosonic contribution to RG evolution is 
now restrained, absolute stability demands tan$\beta$ $\geq$ 3.0. For $M_t$ = 175 GeV however,
lower bound on tan$\beta$ for both stability as well as metastability goes up, stability completely
ruled out for tan$\beta$ $\leq$ 5.0 for instance. Thus, for $M_t$ = 175 GeV, the proportion of metastable model points is higher compared to what is seen for $M_t$ = 175 GeV. The robustness of
this claim is verified by the plots for masses $\simeq$ 1000 GeV, which depict the same qualitative behaviour. Having pointed out the crucial role played by the parameter tan$\beta$, we close this
section here.

\section{Summary and outlook}\label{Conclusions} 

This work highlights the possibility of a metastable EW vacuum in a popular 2HDM
framework. While the occurrence of a panic vacuum in a 2HDM is by and large disfavored
by the latest data from the LHC, we observe that a global minimum at high scales is indeed
possible, iff RG effects are incorporated into the scalar potential. This is found
to happen in the direction of the scalar field $h_2$, because $\l_2$ can be driven
to negative values at high scales . We have reported our findings
in context of a type II 2HDM. 

We remark that it is the relative strengths of the fermionic and bosonic contributions in the RG improved potential that seals the fate of the EW vacuum where we currently reside. The introduction of additional
bosonic degrees of freedom further introduces a tension between vacuum stability on the one hand, and, high-scale perturbativity on the other. This tension can be responsible for substantial constraints on the
parameter space.

In a 2HDM, the
strength of the fermionic contribution is controlled by not only the top quark pole mass, but also
tan$\beta$. Based on the
results of this work, one would always expect a metastable model point in the vicinity of a point
allowing for absolute stability. However, tan$\beta$ picks up a lower bound from the requirement
of metastability, which is tightened when one demands absolute stability of the EW vacuum. The sensitivity of the results to the top pole mass has also been emphasized.

A pertinent extension would be to include finite temperature corrections
to the 2HDM scalar potential and, study its impact on vacuum stability.

\section{Acknowledgements}

We thank Ashoke Sen for useful discussions. This work was partially supported by funding available from the Department of Atomic
Energy, Government of India, for the Regional Centre for Accelerator-based Particle Physics (RE-
CAPP), Harish-Chandra Research Institute.

%%%%%%%%%%%%%%%%%   References %%%%%%%%%%%%%%%%%%%%%%%%%%%%%%%%%%%%
\bibliographystyle{JHEP}
\bibliography{ref.bib}        
\end{document}